\documentclass[namedreferences]{SolarPhysics}

\usepackage[optionalrh,solaenum]{spr-sola-addons} 
\usepackage[dvips]{graphicx}
\usepackage{url}                         

\newcommand{\degrees}{\mbox{$^{\circ}$}}		
\newcommand{\arcsec}{\mbox{$^{\prime\prime}$}}		

\newcommand{\eg}{{\it e.g.}}
\newcommand{\ie}{{\it i.e.}}
\newcommand{\vs}{{\it vs.}}
 \usepackage{color}           
 \definecolor{DarkGreen}{rgb}{0.0,0.45,0.0}  

\begin{document}
\bibliographystyle{spr-mp-sola-cnd}
\begin{article}

\begin{opening}

\title{3D Reconstruction of a Rotating Erupting Prominence}

\author{W. T. \surname{Thompson}$^{1}$\sep
	B. \surname{Kliem}$^{2,3,4}$\sep
	T. T\"{o}r\"{o}k$^{5,6}$}

\institute{$^{1}$ Adnet Systems Inc., NASA Goddard Space Flight Center, Code
	671, Greenbelt, MD 20771, USA\\
	email: \url{William.T.Thompson@nasa.gov}\\
	$^{2}$ Insitut f\"{u}r Physik und Astronomie, Universit\"{a}t Potsdam,
	Potsdam 14476, Germany\\
	$^{3}$ Mullard Space Science Laboratory, University College London,
	Holmbury St. Mary, Dorking, Surrey RH5 6NT, UK\\
	$^{4}$ Part of this work was done while the author visited the
               Space Science Division of the Naval Research Laboratory,
               Washington, DC 20375, USA\\
	$^{5}$ LESIA, Observatoire de Paris, CNRS, UPMC, Universit\'e Paris
               Diderot, 5 place Jules Janssen, 92190 Meudon, France\\
        $^{6}$ now at Predictive Science, Inc., 9990 Mesa Rim Road, Ste.\ 170,
               San Diego, CA 92121, USA}

\begin{abstract}
A bright prominence associated with a coronal mass ejection (CME) was seen erupting
from the Sun on 9~April 2008.  This prominence was tracked by
both the {\it Solar Terrestrial Relations Observatory} (STEREO)
EUVI and COR1 telescopes, and was seen to rotate about the line of sight as it
erupted; therefore, the event has been nicknamed the ``Cartwheel CME.''
The threads of the prominence in the core of the CME quite clearly
indicate the structure of a weakly to moderately twisted flux rope 
throughout the field of view, up to heliocentric heights of 4~solar radii.
Although the STEREO separation was 48{\degrees}, it was possible to
match some sharp features in the later part of the eruption as seen in the
304~{\AA} line in EUVI and in the H$\alpha$-sensitive bandpass of
COR1 by both STEREO {\it Ahead} and {\it Behind}.  These
features could then be traced out in three-dimensional space, and reprojected
into a view in which the eruption is directed towards the observer.  The
reconstructed view shows that the alignment of the prominence to the
vertical axis rotates as it
rises up to a leading-edge height of $\approx\!2.5$ solar radii, and then
remains approximately constant.  The alignment at 2.5 solar
radii differs by about 115{\degrees} from the original filament
orientation inferred from H$\alpha$ and EUV data,
and the height profile of the rotation, obtained here for the first time,
shows that two thirds of the total rotation is reached within
$\approx\!0.5$~solar radii above the photosphere. These features
are well reproduced by numerical simulations of an unstable moderately
twisted flux rope embedded in external flux with a relatively strong
shear field component.
\end{abstract}


\end{opening}

\section{Introduction}

The bandpass of the inner coronagraph (COR1) telescopes on the two {\it Solar
Terrestrial Relations Observatory} (STEREO) spacecraft runs from 650--670~nm
\cite{howard:2008}.
This range was selected to include the hydrogen H{$\alpha$} line at 656~nm, so
as to be sensitive to erupting prominences associated with coronal mass
ejections (CMEs).  A particularly bright prominence eruption was observed by
both COR1 telescopes on 9~April 2008.  Although the relative contributions
of H{$\alpha$} emissions and Thomson-scattered light cannot be derived from the
COR1 observations alone, it is presumed that H{$\alpha$} is a significant
contributor for the extremely bright parts of the prominence early in the
eruption.  This CME and prominence eruption was
also observed by the STEREO {\it Extreme Ultraviolet Imager} (EUVI) telescopes
\cite{howard:2008}
before entering the COR1 field of view, as well as by TRACE at 171~{\AA}
\cite{handy:1999}, and
by the {\it Hinode} XRT telescope \cite{golub:2007} and EIS spectrometer \cite{culhane:2007}.
\citeauthor{Landi&al2010} (2010) analyze this event, combining
{\it Hinode}/XRT and EIS data, together with {\it Solar and Heliospheric
Observatory} (SOHO)/UVCS/EIT/LASCO and
STEREO/EUVI/COR1/COR2 images, to characterize the thermal properties of the
ejected plasma, and constrain the heating rate.
The post-CME current sheet following this event is analyzed by
\inlinecite{savage:2010} and \inlinecite{ko:2010}.
\inlinecite{Patsourakos&Vourlidas2011} determine the geometrical
parameters of the CME and post-CME current sheet in the field of view of
the outer coronagraph (COR2) on STEREO. 
This event has become known as the ``Cartwheel CME'' because of the highly
visible rotation around the line of sight seen during the initial stages of the
eruption.  Rotation about the vertical direction can also be
observed through triangulation as the prominence material rises through the
EUVI and COR1 height ranges. The plane of sky motions which give this event
its ``Cartwheel'' name is actually a combination of the rotation about the
vertical axis together with a deflection of the prominence in latitude and
longitude during the initial rise phase. A rotation around the axis
of the prominence, known as ``roll effect'' \cite{martin:2003}, may have
contributed as well. Further rotation about the vertical direction
is indicated by the CME orientation in the COR2 data.  This paper presents an
exploration of the rotation about the vertical direction in the EUVI and COR1
height ranges,
using triangulation from the two STEREO viewpoints. The height profile
of the rotation angle for the erupting prominence is obtained, which
could not be done prior to the STEREO mission.

The separation of the two STEREO spacecraft at this time was 48{\degrees}.
Co-identification of features in both views is quite difficult with such a
large separation.  However, it was possible to match some sharp features in the
later part of the eruption as seen by EUVI in the He~{\sc ii} line at
304~{\AA}.  Features were also tracked in white light (with a possible
H{$\alpha$} component) as seen by COR1,
although not necessarily the same features seen at 304~{\AA}.

The derived 
{height profile of prominence rotation is compared with the
orientation of the  polarity inversion line of the radial field
component (PIL) at various heights, to check whether the rotation is
simply the result of alignment with the PIL. The radial field
component is obtained from a potential-field source-surface (PFSS)
extrapolation. Additionally, the rotation} 
and rise profiles of the prominence, 
as well as the STEREO images, are compared with the
corresponding data obtained in a series of CME simulations in a
companion paper
(\opencite{Kliem&al2011};
in the following Paper~II),
in order to {explore a broader range of possible origins}
of the large total rotation observed. The
{comparisons suggest} that the major part of the rotation was due to the
presence of a shear field component of considerable strength in the
source region \cite{isenberg:2007} and that the weak helical kink
instability of a moderately twisted flux rope \cite{torok:2004}
contributed as well. Field line plots of the best matching case are
included here to demonstrate the correspondence with the observed
overall shape and helical threads of the rising prominence.

\section{Data Analysis}
\label{data_analysis}

\subsection{Source Region}
\label{source_region}

\begin{figure}
\centering
\includegraphics[width=0.67\textwidth]{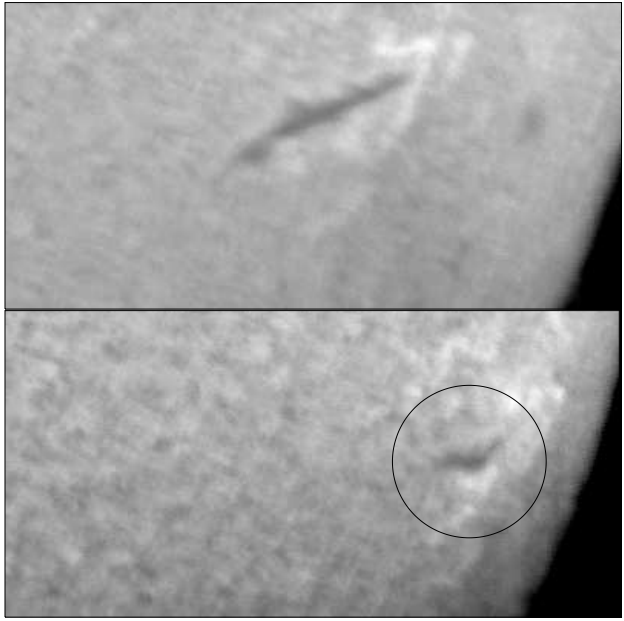}
\caption[]
{H{$\alpha$} image of the filament as seen on the disk on 5~April 2008,
01:51~UT, before the first eruption on that day, and on 6~April,
02:44~UT (circled), reformed after the two eruptions on 5~April
(from the Yunnan Astronomical Observatory in Kunming, China).}
\label{filament}
\end{figure}

\begin{figure}
\centering
\includegraphics[width=0.8\textwidth]{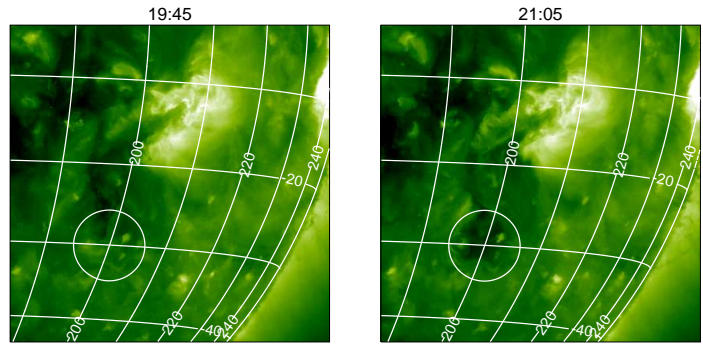}
\caption[]
{STEREO {\it Ahead} EUVI 195~{\AA} images of AR~10989 before and
 after the second filament eruption on 5~April 2008, displaying an area
 of 640{\arcsec} on a side. The image on the left at 19:45:30~UT shows
 the full extent of the filament on the souteastern end at the onset
 of the eruption (up to the middle of the circle), and the image
 on the right at 21:05:30~UT (with the circle plotted at the same
 position) shows that a dimming has developed at this location.
 Therefore, the erupting flux should be rooted in this area. Comparison
 with Figure~\ref{filament} shows that only the upper branch of the
 filament is visible in H$\alpha$.
 Overplotted grid lines represent Carrington longitude and latitude.
}
\label{dimmings}
\end{figure}

\begin{figure}
\centering
\includegraphics[width=0.5\textwidth]{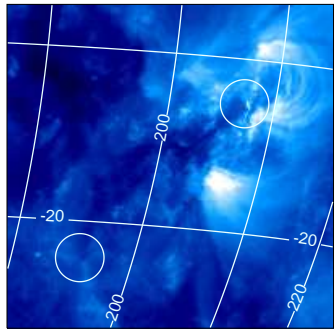}
\caption[]
{STEREO {\it Ahead} EUVI 171~{\AA} image of the reformed filament
 on 6~April 2008, 15:01~UT. The area shown is 320{\arcsec} on a side.
 The circles mark the ends of the filament material visible at this
 wavelength. Their distance is $\approx175$~Mm. The main body of
 the filament as seen in this line is oriented at an inclination of
 $\approx\!26\degrees$ to the east-west direction on the Sun.
}
\label{reformed_filament}
\end{figure}

\begin{figure}
\centering
\includegraphics[width=0.7\textwidth]{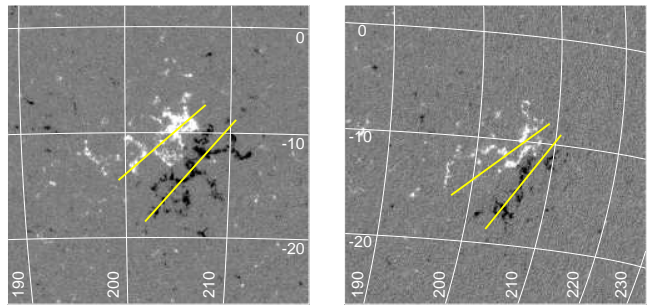}
\caption[]
{Magnetogram of AR~10989 on 31~March \textit{(left)} and on 4~April 2008
 \textit{(right)} taken by SOHO/MDI. The overplotted bars indicate rough
 estimates of the distance between the center of each polarity (in a
 center-of-gravity sense), which is an important input parameter for the
 numerical modeling of the event in Paper~II.}
\label{mdi}
\end{figure}

The eruption on 9~April 2008 occurred in the remnants of NOAA active region
(AR) 10989, located about 23{\degrees}
behind the west limb as seen from Earth, and had an onset time near 09~UT. The
filament had been visible nearly throughout the disk passage of the
active region, which was in its decaying stage and spotless after 31~March.
The filament erupted twice on 5 April 2008, reforming afterwards.  These
eruptions were visable on disk to EUVI on STEREO {\it Ahead}.  Both eruptions
appeared to propagate inclined from the radial direction toward higher southern
latitudes, similar to the eruption on 9~April.  The inclination made it
difficult to discern rotational motions
based on images from only a single viewpoint. However, a
stereoscopic reconstruction of the second eruption revealed a considerable
rotation of about 90{\degrees} \cite{bi:2011}.  Interestingly, the rotation
was in the clockwise direction, opposite to that of the eruption on 9~April.

As with most filaments, it is difficult to discern the magnetic connections at
its ends.  However, both eruptions on 5~April produced a dimming
and endpoint brightenings in the EUV \cite{wang:2009} near the southeastern
end of the
filament, which were located in negative polarity, the polarity which dominated
the southwestern side of the filament channel, so that the filament was dextral
according to the classification of \citeauthor{martin:1998} (1998).  This is
further supported by the right-bearing orientation of the filament barbs (see
Figure~\ref{filament}).  Dextral filaments are embedded in field of left
handed chirality, as indicated by the typical skew of the overlying arcade of coronal
loops \cite{martin:1998}.  If the assumption of flux rope topology holds for
the filament, then the filament itself is also threaded by left-handed field.
Otherwise, it may have the opposite chirality (see \opencite{martin:1997};
\opencite{ruzmaikin:2003}; \opencite{Muglach&al2009}).  If a filament
in dominantly left-handed helical field rotates upon eruption, the direction of
rotation is generally found to be in the counter-clockwise direction, as
expected from helicity conservation \cite{rust:2005,green:2007}.

{Note that the dextral chirality of the filament is an exception to the usual
hemispheric rule \cite{martin:1994}.  However, \citeauthor{pevtsov:2003} (2003)
show that 16--25\% of filaments do not follow the hemispheric rule, and that
the rule is weaker for filaments in active regions.}

On 5~April, the filament was oriented at an inclination of
$\approx24\degrees$ to the east-west
direction on the Sun and had a length of $\approx 135$~Mm in the image shown in
Figure~\ref{filament} (top).  However, the dimmings and endpoint
brightenings formed in the eruptions later on the same day suggest that the
true extent of the flux in the filament channel was larger by about one third,
$\approx 175$~Mm, see Figure~\ref{dimmings}.

Due to foreshortening, the H$\alpha$ images on the following day show the
filament, its magnetic connections and barbs far less clearly, but it can be
seen that the filament reformed in a similar location
%
%
(Figure~\ref{filament}, bottom).
The better perspective of the EUVI-{\it Ahead} images indicates that the
main body of the
reformed filament was nearly straight, oriented at a tilt angle
of $\approx26\degrees$ to the east-west direction,
and extended across a similar
length as on 5~April (Figure~\ref{reformed_filament}). This is most
clear if the 171~{\AA} and 304~{\AA} images are viewed in animated
format. As far as the increasing foreshortening allows a judgment, the
filament appeared to keep its orientation and shape through the
subsequent days until a new eruption launched the Cartwheel CME.

Since the numerical modeling in Paper~II reveals a strong influence of
the distance between the polarities in the source region on the
rotation of the ejected flux rope, we attempt to estimate this parameter.
Figure~\ref{mdi} shows magnetograms of the region from the
SOHO Michelson Doppler Imager (MDI) on 31~March and 4~April 2008. The
overplotted bars run through the middle of each polarity in a
one-dimensional center-of-gravity sense, providing an indication of
the distance between the main polarities. This value increases from
$\sim\!40$~Mm to a range of $\sim\!(40\mbox{--}75)$~Mm over the four-day
time span.  A range
twice as wide may have been characteristic of the configuration after
the similar time span to the eruption on 9~April. It appears
impossible to estimate a single relevant value. We will adopt a
distance of 90~Mm as a base value in most simulations in Paper~II, but
also study the influence of its variation. This base value lies in the
middle of the estimated range, so it is consistent with the fact that
much of the rotation occurred in the course of the radial ascent above
the south-east edge of the remnant active region, which is
characterized by a considerable distance between the polarities (see
Section~\ref{ss:measurements}).

{The magnetograms in Figure~\ref{mdi} show an overall displacement
of the two polarities in the direction of the PIL. While the negative
flux is more or less uniformly distributed along the filament channel,
the positive flux is more concentrated near the northern end. This
results in a net shear field component pointing along the filament
channel in the southeastward direction. Coronal field lines that arch
above the filament thus possess a skew corresponding to the upper part
of a left-handed helix, in agreement with the implication made above
that the filament was dextral.}

We have also estimated the height of the prominence from data taken by
the EUV Imaging Telescope (EIT)
\cite{delaboudiniere:1995} onboard SOHO on 8~April, when the prominence
was close to the limb.  The material seen at that time in the 304~{\AA}
line extended up to
a height of \mbox{$\approx$ 0.04~$R_\odot$} above the photosphere, and the
absorbing material seen at 171 and 195~{\AA} extended up to \mbox{$\approx$
0.033~$R_\odot$}.  From EUVI 304~{\AA} data about an hour before the eruption,
we estimate the height to lie in the range \mbox{$\approx$ 0.05--0.06
$R_\odot$}, so it is clear that the prominence experienced a slow rise before
the eruption.

\begin{figure}
\centering
\includegraphics[width=0.8\textwidth]{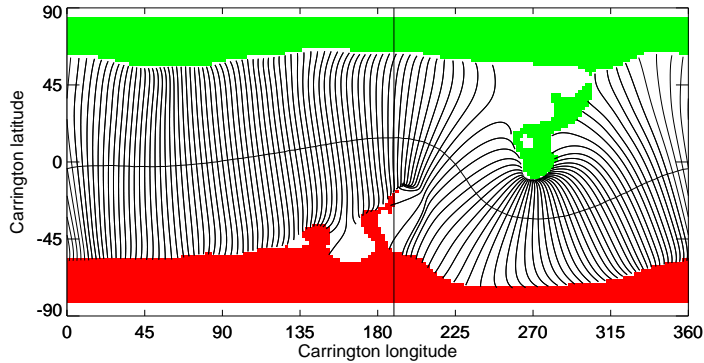}
\caption[]
{Location of the heliospheric current sheet for Carrington Rotation~2068
 from a potential-field source-surface extrapolation. The current
 sheet runs nearly east-west above the estimated CME direction of propagation
 as indicated by the vertical bar.
 The green and red regions represent open field regions of negative and
 positive polarity respectively.}
\label{pfss}
\end{figure}

Motions along the prominence could be seen in the 171~{\AA} channel
of EUVI-{\it Ahead} on 9~April from about 08:20~UT onward. The first upward motions
of prominence material along the CME path can be discerned by comparing the
images at 08:48:30 and 08:51~UT. \inlinecite{Landi&al2010} give a rather
conservative estimate of the CME start time, 09:10~UT, but the 171~{\AA}
images, taken at 2.5~min cadence, show that most of the prominence body
visible to EUVI-{\it Ahead} was already moving by 08:53:30~UT, the start
time quoted in \inlinecite{savage:2010}. Since the event
commenced in the absence of a strong perturbation (no signs of a
significant brightening, of a jet, or of perturbations resulting from
nearby activity were seen), it must have developed from a small
perturbation when the configuration was near the boundary between stable
and unstable states. In such cases the initial motion of the unstable
flux is expected to behave exponentially,
which requires some time to develop to a level
that causes visible changes. Rapid changes in the vertical
position were first seen between the EUVI-{\it Ahead} images at 08:48:30
and 08:51~UT. Therefore, the actual start time should lie before
08:51~UT.

A helmet streamer, best seen in the COR2-{\it Ahead} images, extended above
{AR~10989 in the plane-of-sky projection}.
In order to estimate the orientation of the
heliospheric current sheet above the streamer, {and whether it was
magnetically connected with the PIL in the active region,} we ran a
{PFSS extrapolation} and a
Wang-Sheeley-Arge model for Carrington Rotation~2068 at the
Community Coordinated Modeling Center (CCMC), using a source surface
location at 2.5~solar radii for both models. Figure~\ref{pfss} shows
the {PIL and corresponding field lines originating at the 2.5~$R_\odot$
source surface}
of the PFSS model,
which yields a nearly east-west
orientation of the heliospheric current sheet {at the longitude of
radial CME propagation estimated in Section~\ref{ss:measurements}.}
The Wang-Sheeley-Arge model gives the same result. 
{It is clearly seen that the field at large heights above the active
region is dominated by the large-scale field of the Sun, which points
essentially opposite to the shear field component in the active region
at small heights.}

\begin{figure}
\centering
\includegraphics[width=\textwidth]{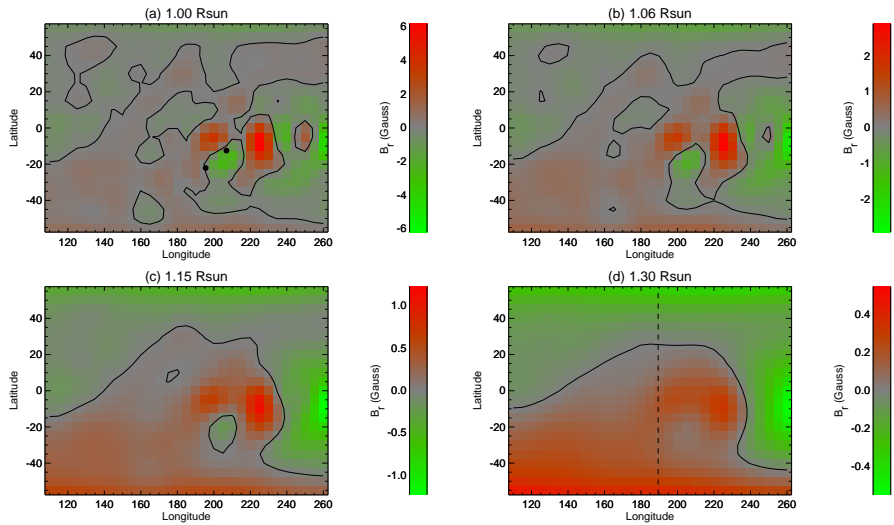}
\caption[]
{Polarity inversion line of the radial field component in the PFSS
model, (a) in the photosphere, (b) at the estimated initial height of
the prominence, (c) at the largest height that shows an
orientation of the PIL above AR~10989 relatively clearly,
and (d) at 1.3~$R_\odot$ where the PIL above AR~10989 is
completely gone.
Dots in panel (a)
mark the end points of the prominence shown in
Figure~\ref{reformed_filament}, and the vertical line in
panel (d) marks the estimated longitude of the CME, as in
Figure~\ref{pfss}.}
\label{pfss2}
\end{figure}

{A similar representation of the radial field component of the PFSS
model at 
several representative heights
is given in Figure~\ref{pfss2}.
The PIL above AR~10989 is part of a magnetic structure which is
distinct from the PIL at the base of the heliospheric current sheet,
and which is seen as a separate ring in the low corona (\eg\ at
1.15~$R_\odot$).
With increasing height, the
ring-shaped PIL disappears (between 1.2 and 1.3 solar radii), and the
PIL associated with the heliospheric current sheet gradually
approaches the equator at the estimated longitude of the CME.}

\subsection{3D Reconstruction}

The SolarSoft routine \verb+scc_measure+ was used to measure the
three-dimensional location of features in the STEREO EUVI and COR1
images.  The action of this program is as follows: The user is presented with
two side-by-side images, one from each of the two STEREO spacecraft.  The
images are selected so that they represent the same observation time, with a
slight offset to account for the difference in light travel time from the Sun.
The user can zoom in on the region of interest in the two images, and adjust
the color table and data range to optimize the appearance of the feature being
measured.  A point is selected on one image with the cursor.  The program
calculates the three-dimensional line of sight represented by this point, and
then overplots the projection of this line onto the image from the other
satellite.  This is known as an epipolar line.  Since both the EUVI and COR1
optics produce a gnomonic projection on the CCD detector, straight lines in
space will always appear as straight lines in the image.  The feature selected
by the user in the first image must appear along the epipolar line drawn by the
program in the second image.  The correct location along this line is selected
by the user, which leads to another line of sight calculation which intersects
the original line of sight.  The intersection of these two lines determines the
three-dimensional (3D) location of the feature.
{In this investigation, we use this technique to find the 3D locations of
multiple points along prominence threads.}

The main difficulty in applying this technique to the prominence eruption is
source confusion.  On 9~April 2008, the two STEREO spacecraft were
separated by 48{\degrees} along the ecliptic plane.  Thus, the appearance seen
by STEREO {\it Ahead} was quite different from that seen by STEREO {\it 
Behind}.  This made it very difficult to locate features which could be
positively identified to be the same in both images.  Another goal was
to identify
features which could be tracked through several frames.  (However, no attempt
was made to match identifications between EUVI and COR1.)  In spite of these
difficulties, it was possible to identify several features which could be
identified in both views, and which could be tracked through several frames.

\subsection{EUVI and COR Measurements}
\label{ss:measurements}

Figures~\ref{euvi1} and~\ref{euvi2} show the tracked features as seen in the
304~{\AA} channel of EUVI-{\it Ahead}.
{The colored lines represent the filamentary features that were measured at
each time step.  Each line has a different color, and a dot at one end, to
better show the relationship with the reprojected graphs along the right side.
The different colors do not necessarily represent the same threads at each time
step.}
The He~{\sc ii} 304~{\AA} channel was chosen
as being most representative of the cool prominence material seen later in
a combination of H{$\alpha$} and Thomson-scattered light by COR1.  Although
the eruption is clearly seen by {\it Ahead} at
9:06~UT, and by {\it Behind} at 9:36~UT, source confusion made it impossible to
track features earlier than 10:06~UT.  Initially, little sense could be made
from an examination of these data viewed in 3D (\eg, through an anaglyph
representation).  However, because the measurements are made in 3D, they can be
reprojected to another viewpoint.  By selecting the proper viewpoint, it's
possible to see aspects of the data that are not evident in the frame in which
the data are taken.  The plots to the right in Figures~\ref{euvi1}
and~\ref{euvi2} show the same data as in the images, but from a viewpoint at
Stonyhurst longtitude 98{\degrees} west (relative to Earth), and 24{\degrees}
south.  It is estimated that from this viewpoint, the CME is traveling straight
toward the observer.

From this perspective, a certain organization can be perceived.  The primary
alignment in the 10:06~UT data is vertical, {{\ie} aligned north-south,} with
the upper prominence threads lying almost directly over the lower threads {in
the frame of the figure}.
However, in the later
images the primary alignment is distinctly sloped, with the upper threads
lying toward the east, and the lower ones toward the west.  This is
particularly well seen in the image at 10:46~UT in Figure~\ref{euvi2}, where
the data display an overall slope of $-50\degrees$.  As one goes through the
time steps, a counter-clockwise rotation can be perceived as one steps from
10:06 to 10:46~UT.

The prominence structure must have rotated through $\approx65\degrees$ from its
original orientation to achieve the nearly {north-south} structure seen at
10:06~UT,
followed by an additional $\approx50\degrees$ to reach the maximum rotation
at 10:46~UT.  Thus, the full amount of rotation experienced by the prominence
in the EUVI height range was on the order of 115{\degrees}.

\begin{figure}
\begin{center}
\includegraphics[width=0.475\textwidth]{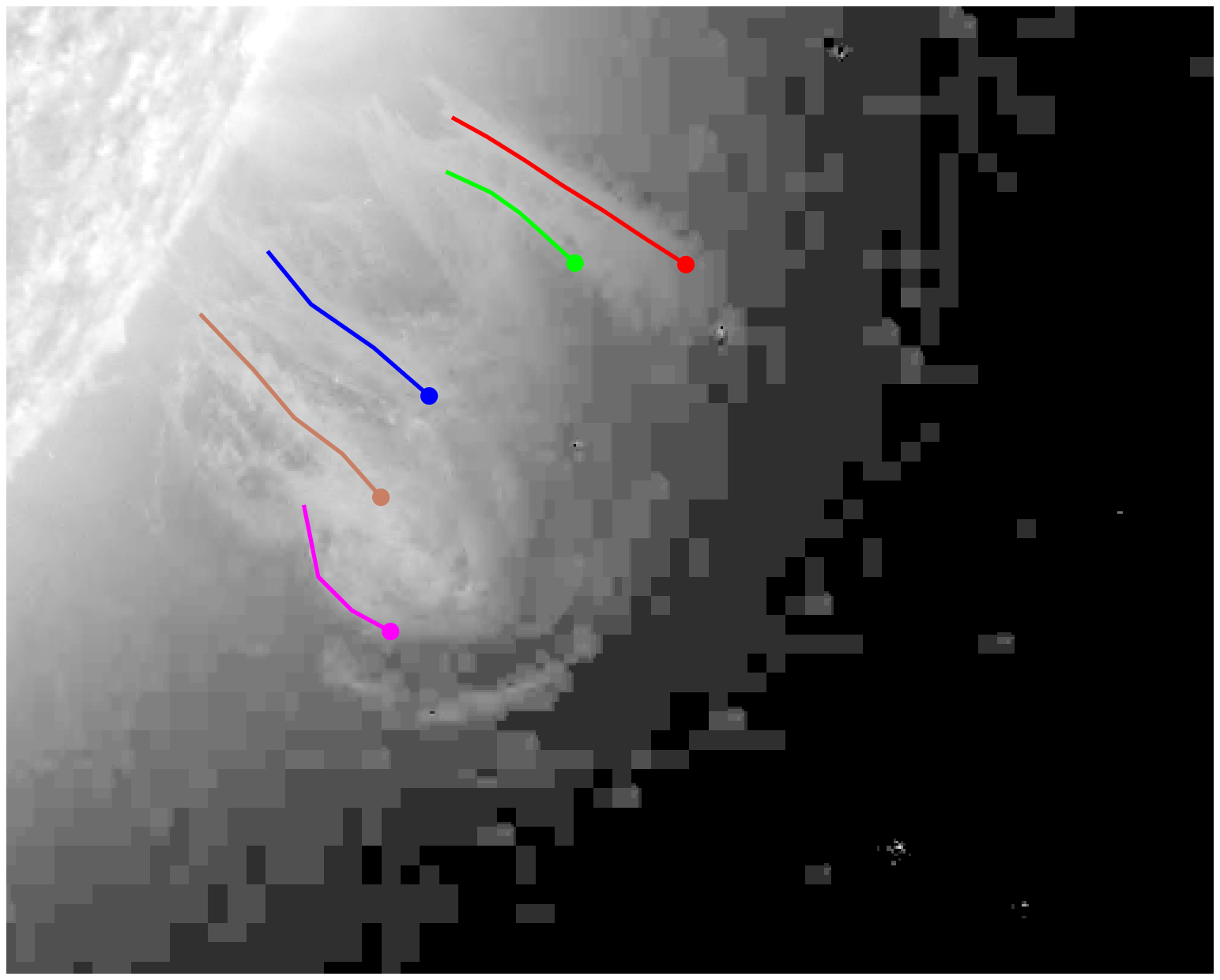}
\includegraphics[width=0.475\textwidth]{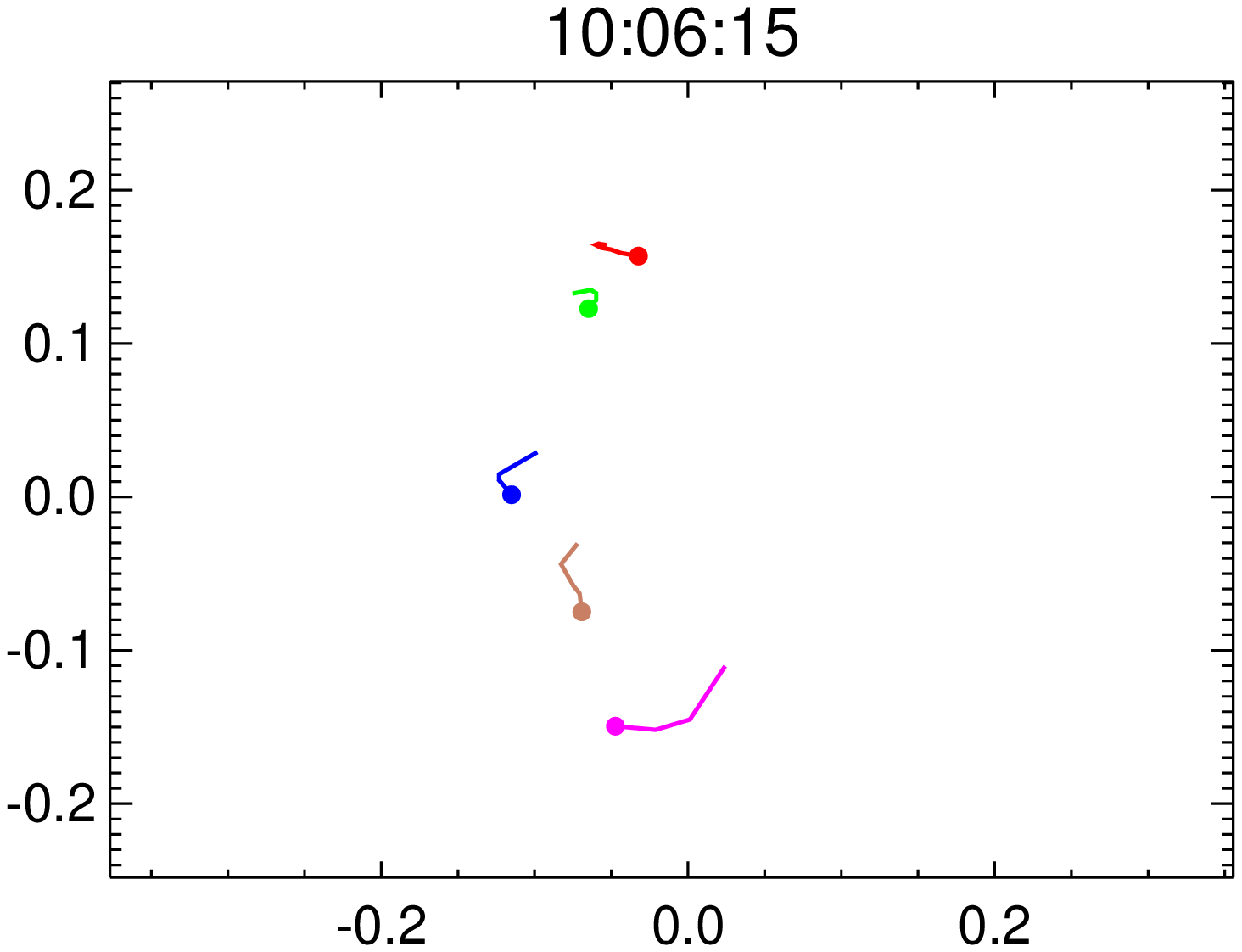}\\
\includegraphics[width=0.475\textwidth]{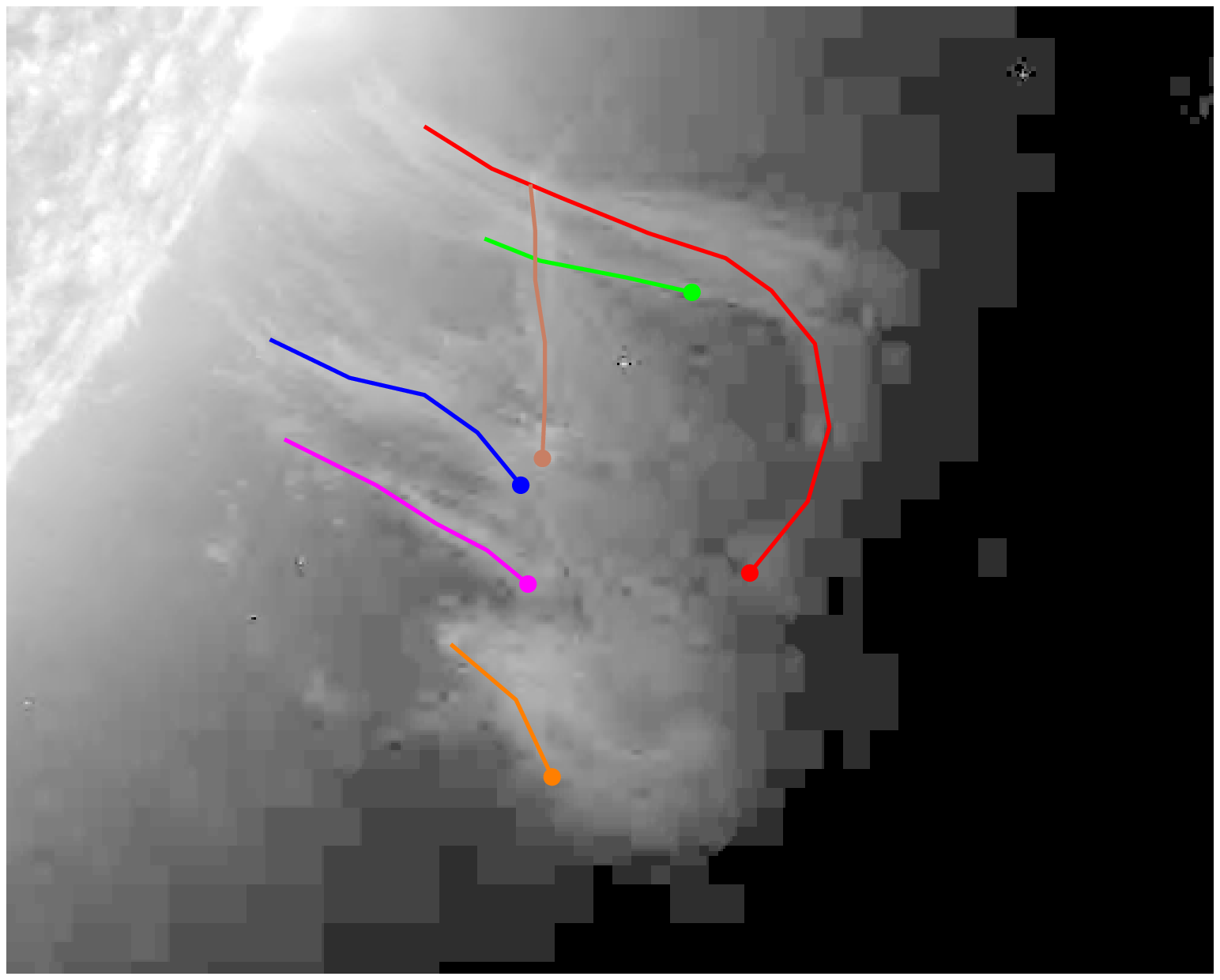}
\includegraphics[width=0.475\textwidth]{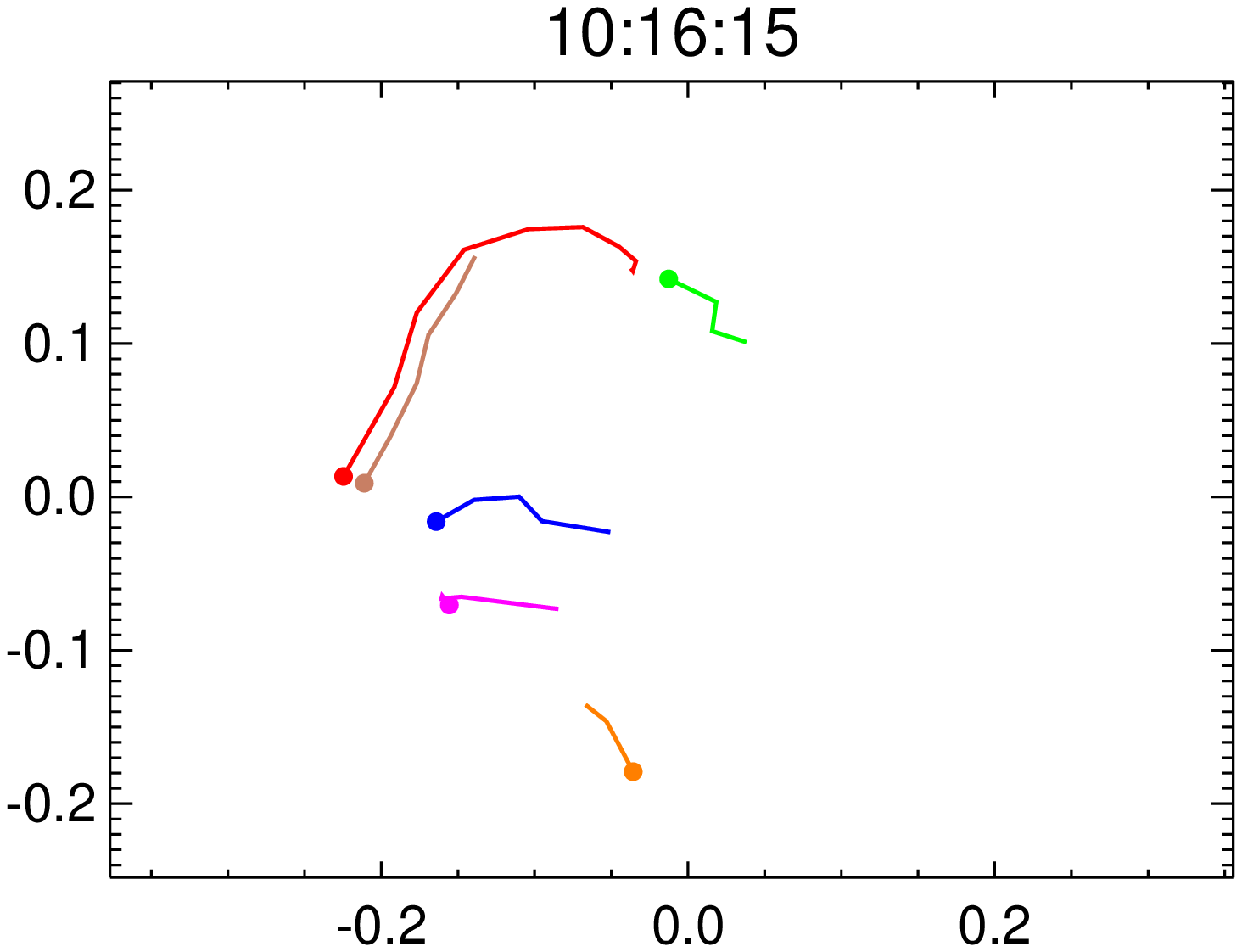}\\
\includegraphics[width=0.475\textwidth]{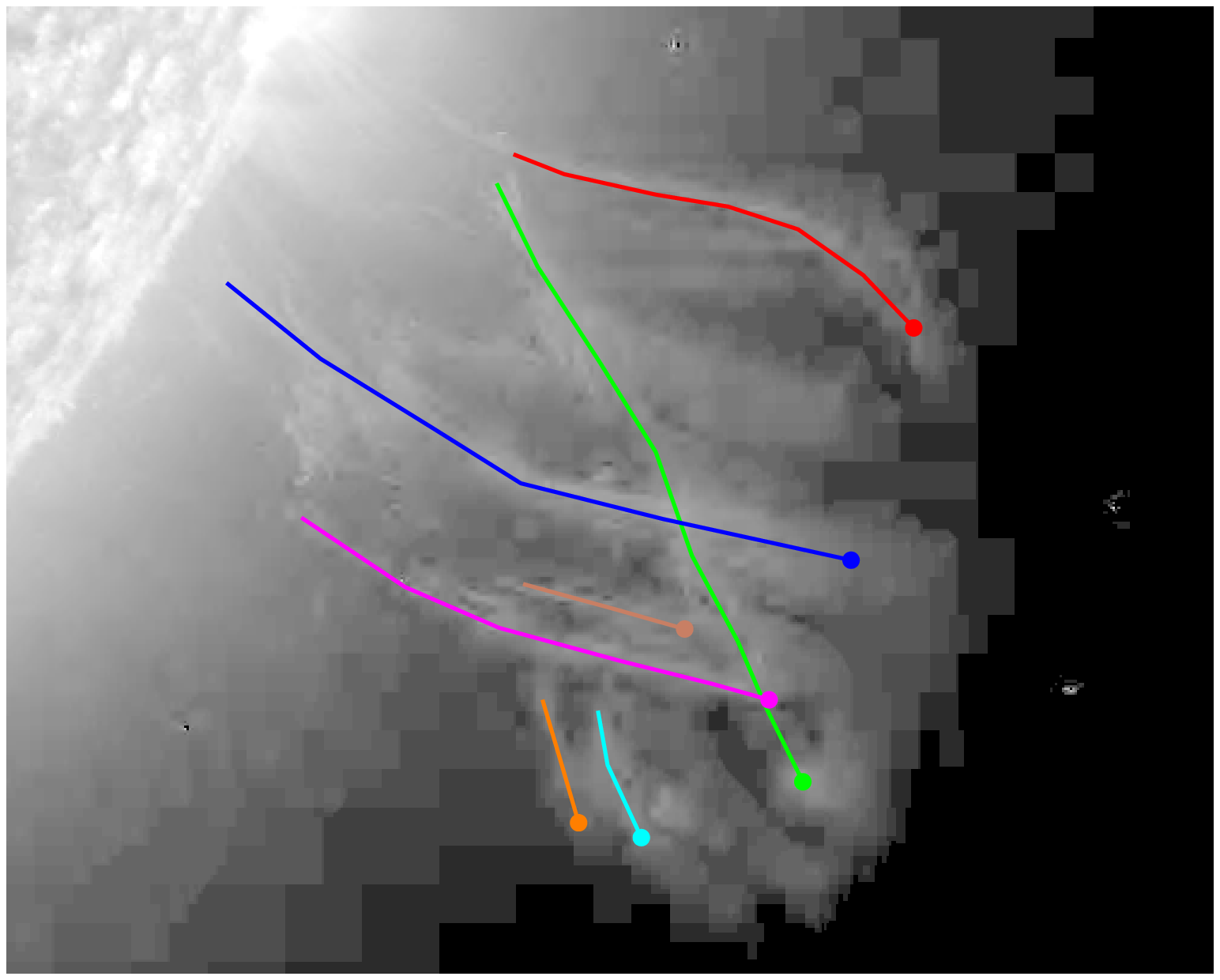}
\includegraphics[width=0.475\textwidth]{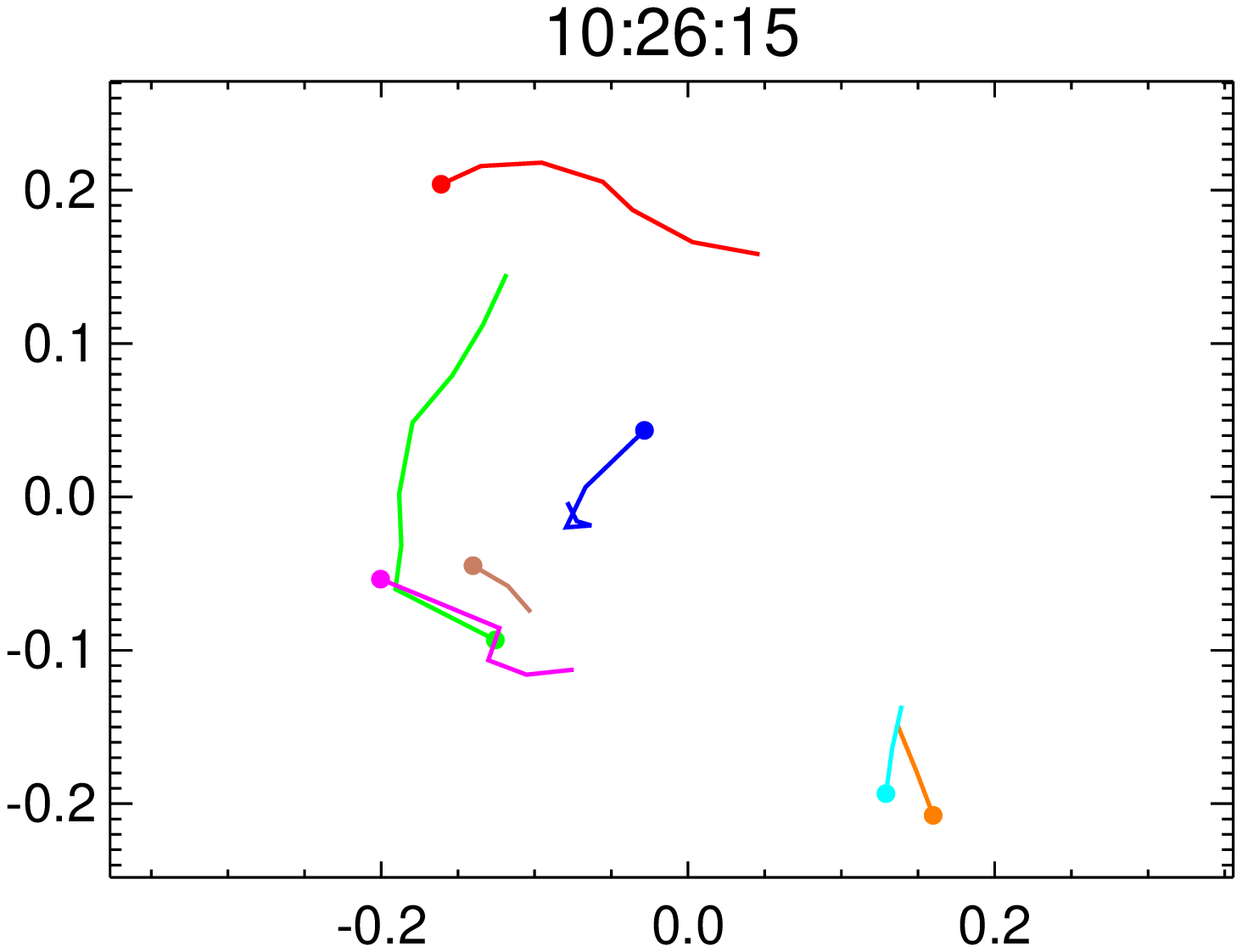}
\end{center}
\caption[]
{Images and plots of the prominence eruption from 10:06 to 10:26~UT, as seen in
the 304~{\AA} channel of the STEREO EUVI telescopes. Subsequent frames are
shown in Figure~\ref{euvi2}.  Along the left are shown the EUVI-{\it Ahead} images
with the tracked features overplotted.
{The lower-right corner of each image is outside the EUVI field of view,
causing the threads to appear truncated in height at 10:26~UT and
subsequent times.} The plots along the right show the same
data reprojected to a viewpoint at Stonyhurst longitude 98{\degrees} west
(relative to Earth) and latitude 24{\degrees} south. Axes are in units of
solar radii.}
\label{euvi1}
\end{figure}

\begin{figure}
\begin{center}
\includegraphics[width=0.475\textwidth]{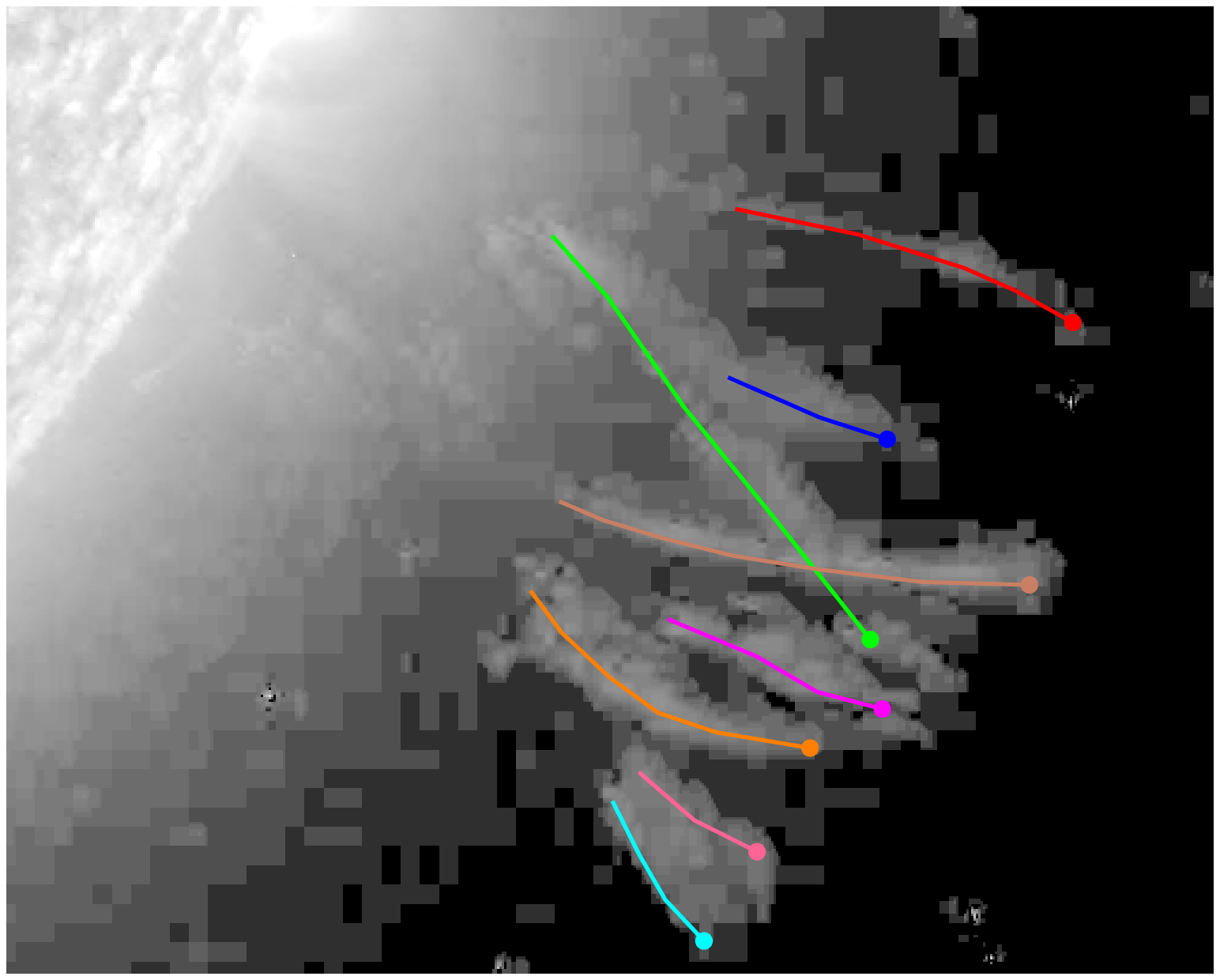}
\includegraphics[width=0.475\textwidth]{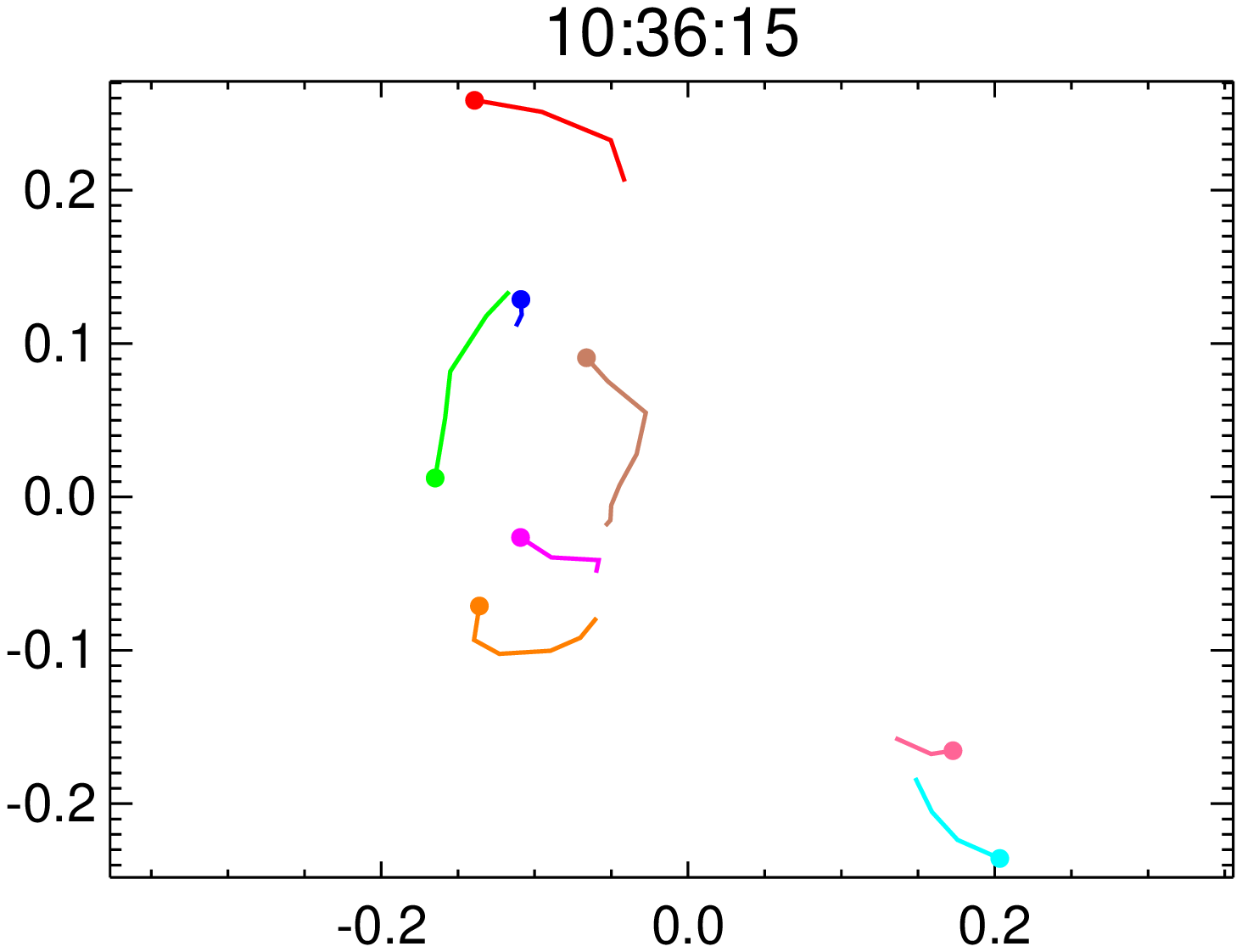}\\
\includegraphics[width=0.475\textwidth]{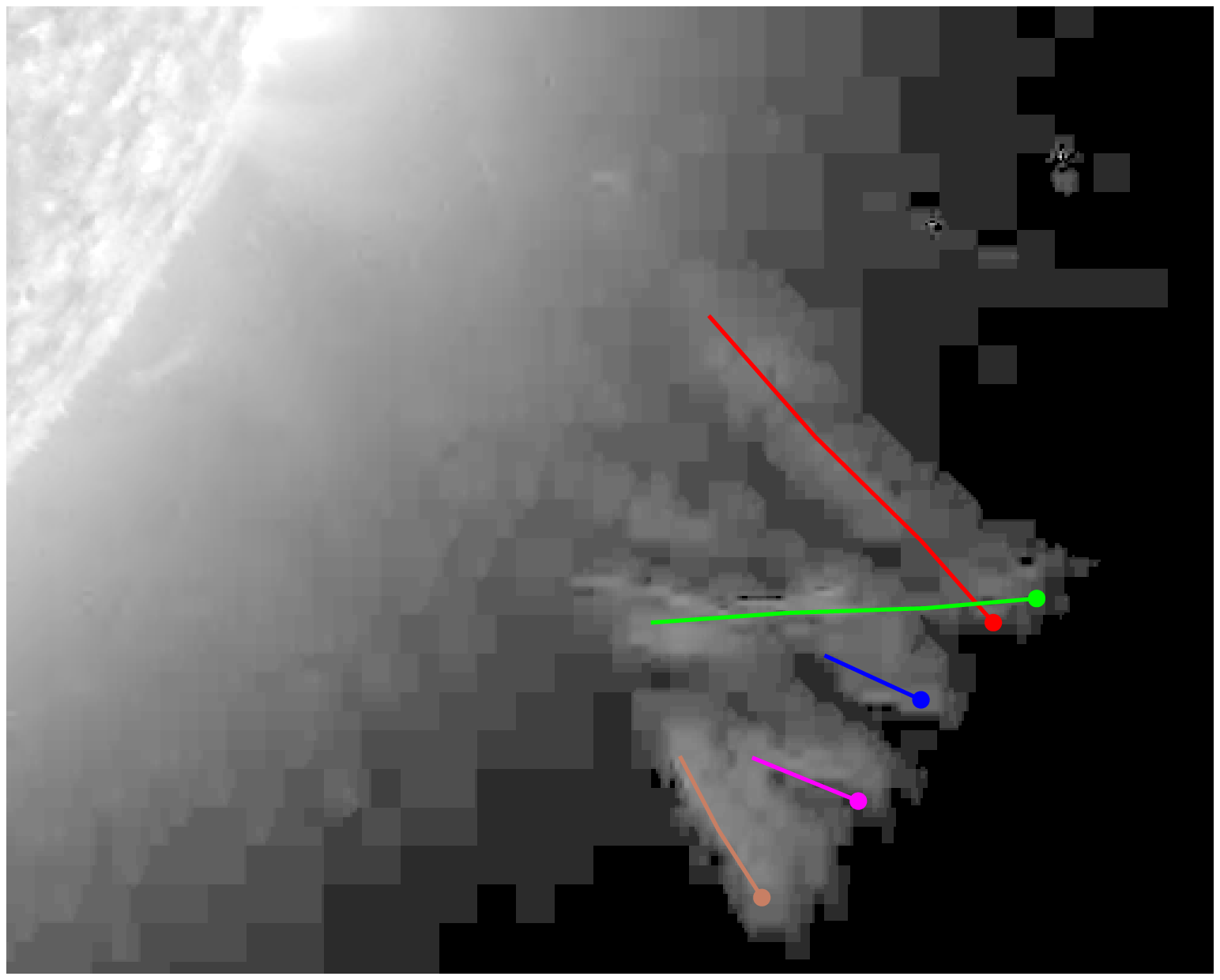}
\includegraphics[width=0.475\textwidth]{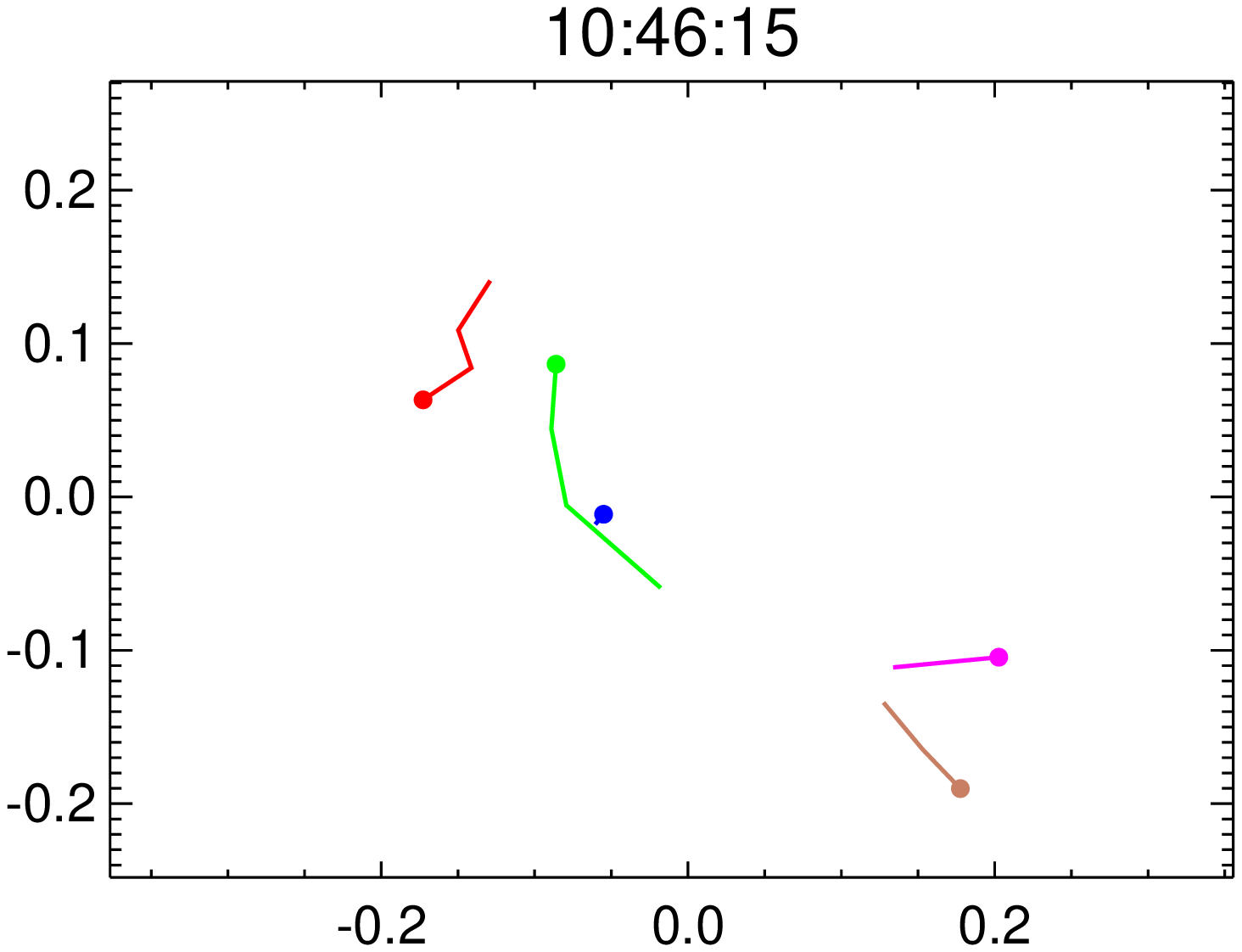}\\
\includegraphics[width=0.475\textwidth]{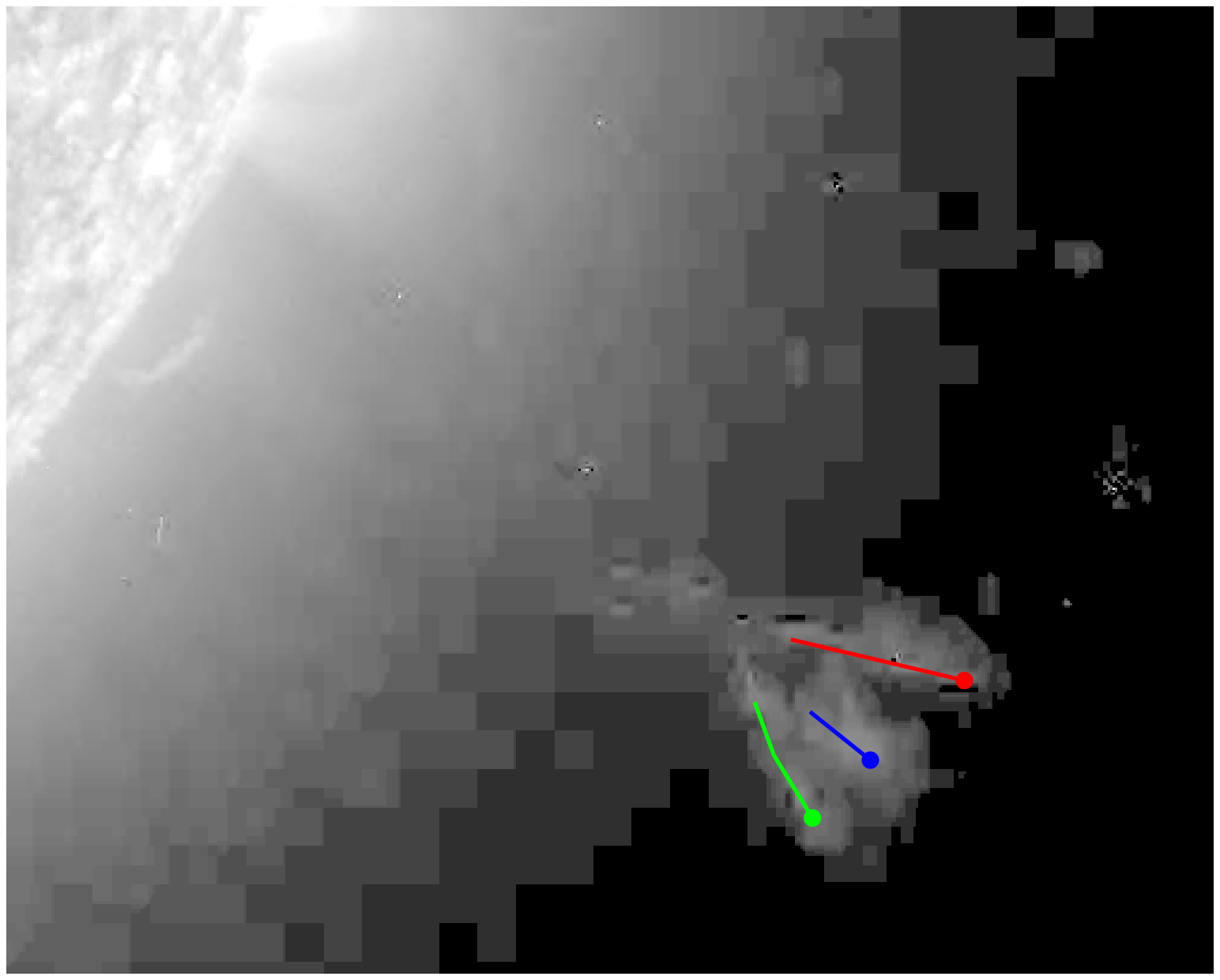}
\includegraphics[width=0.475\textwidth]{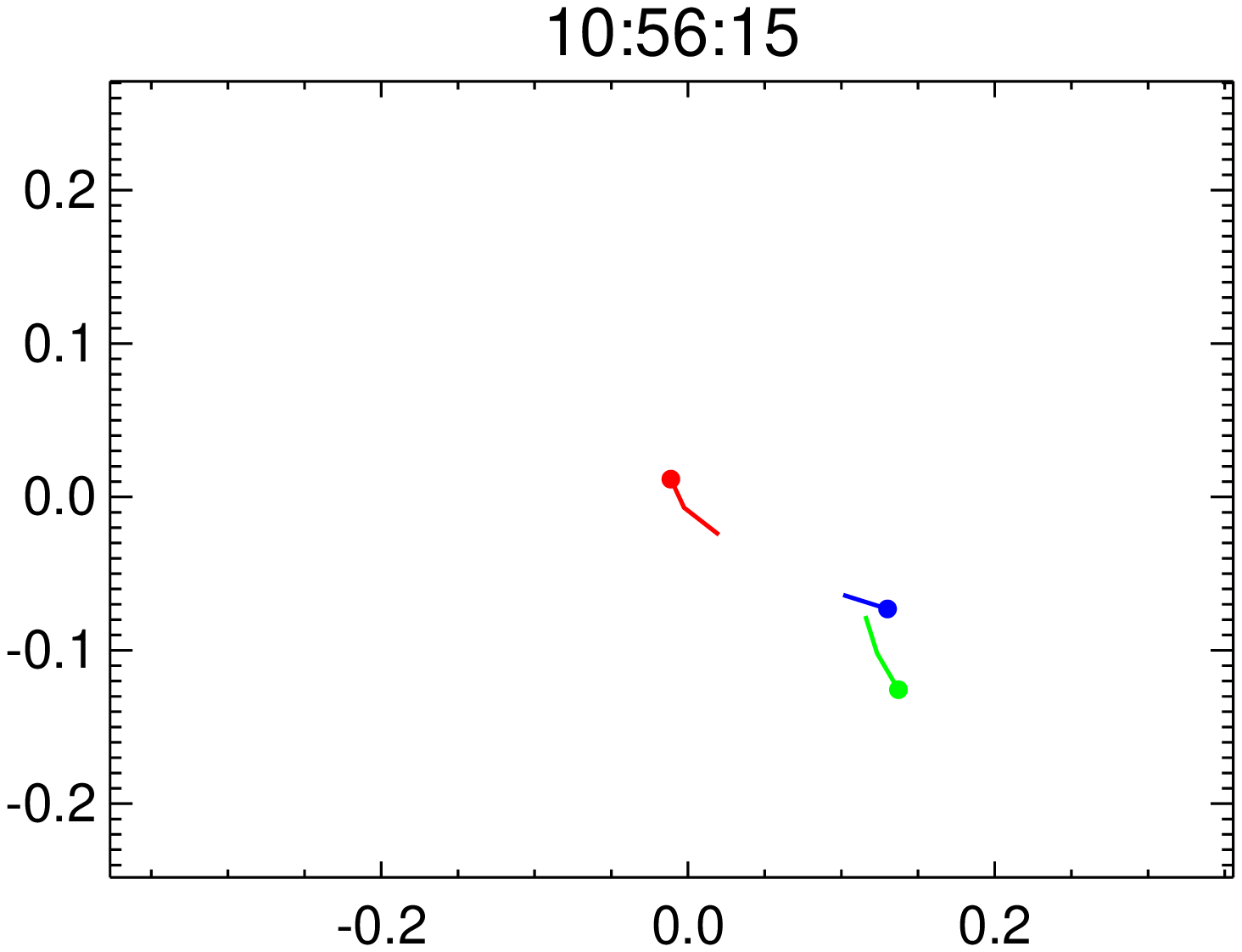}
\end{center}
\caption[]
{Continuation of Figure~\ref{euvi1} for time steps 10:36 to 10:56~UT.}
\label{euvi2}
\end{figure}

\begin{figure}
\begin{center}
\includegraphics[width=0.4\textwidth]{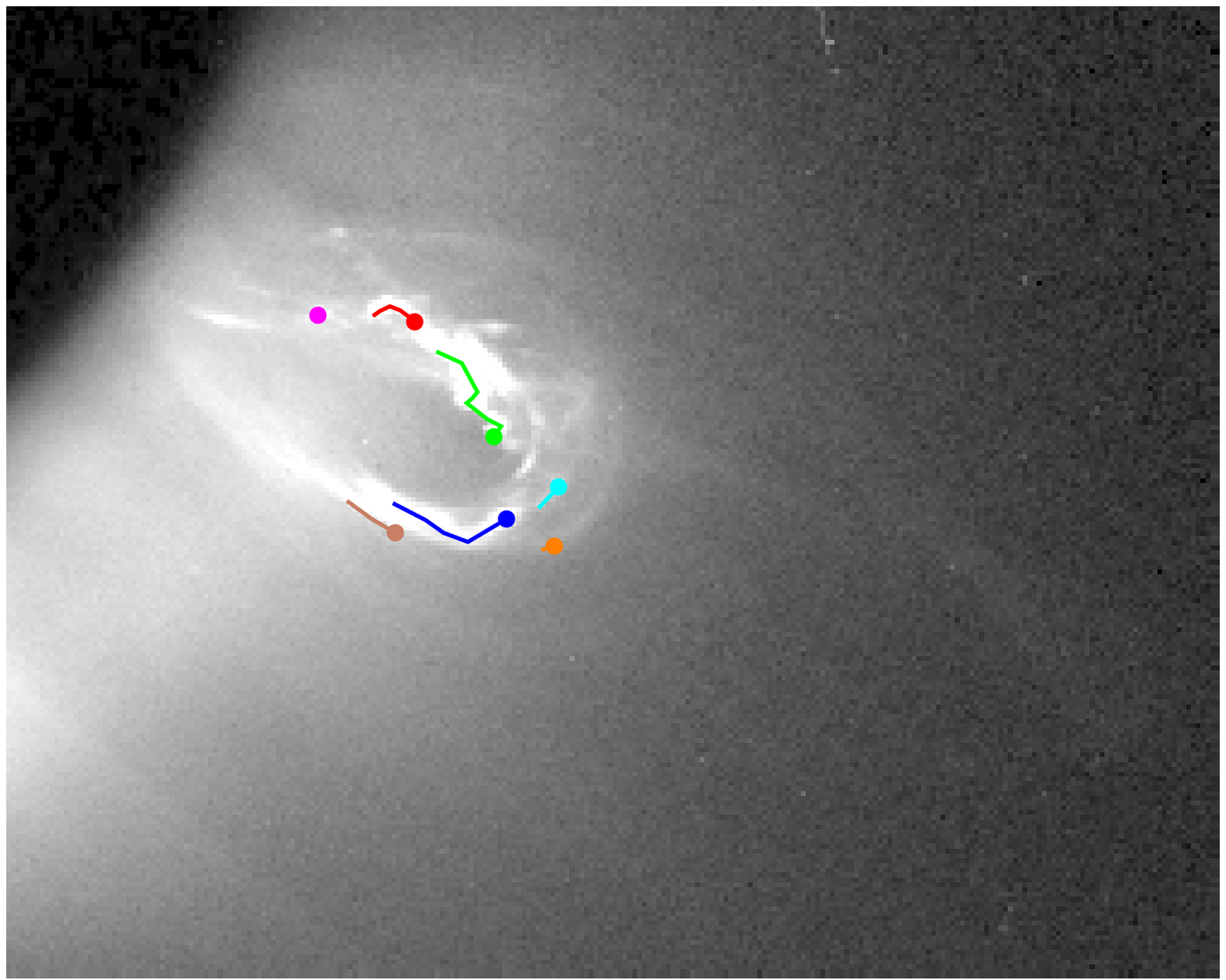}
\includegraphics[width=0.4\textwidth]{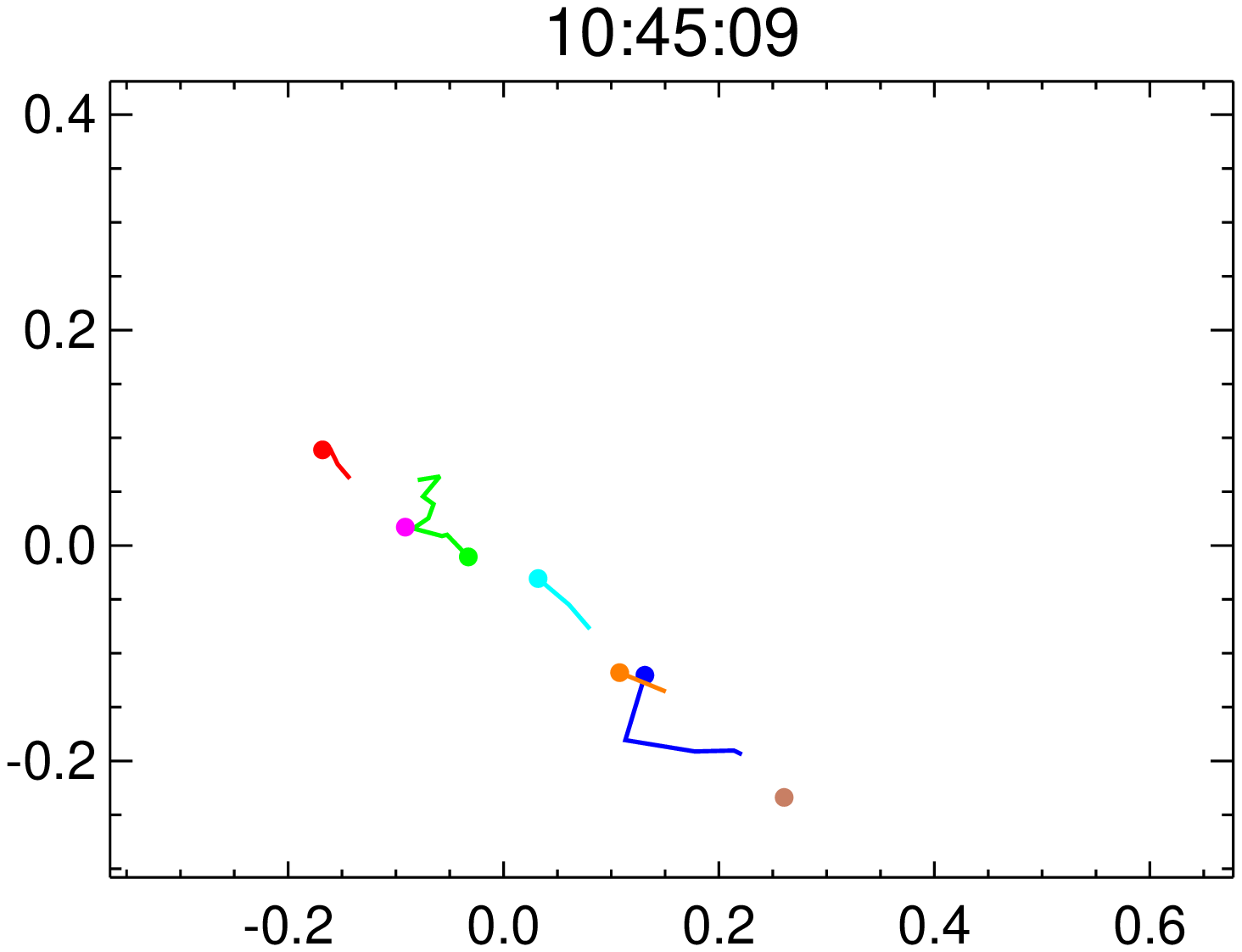}\\
\includegraphics[width=0.4\textwidth]{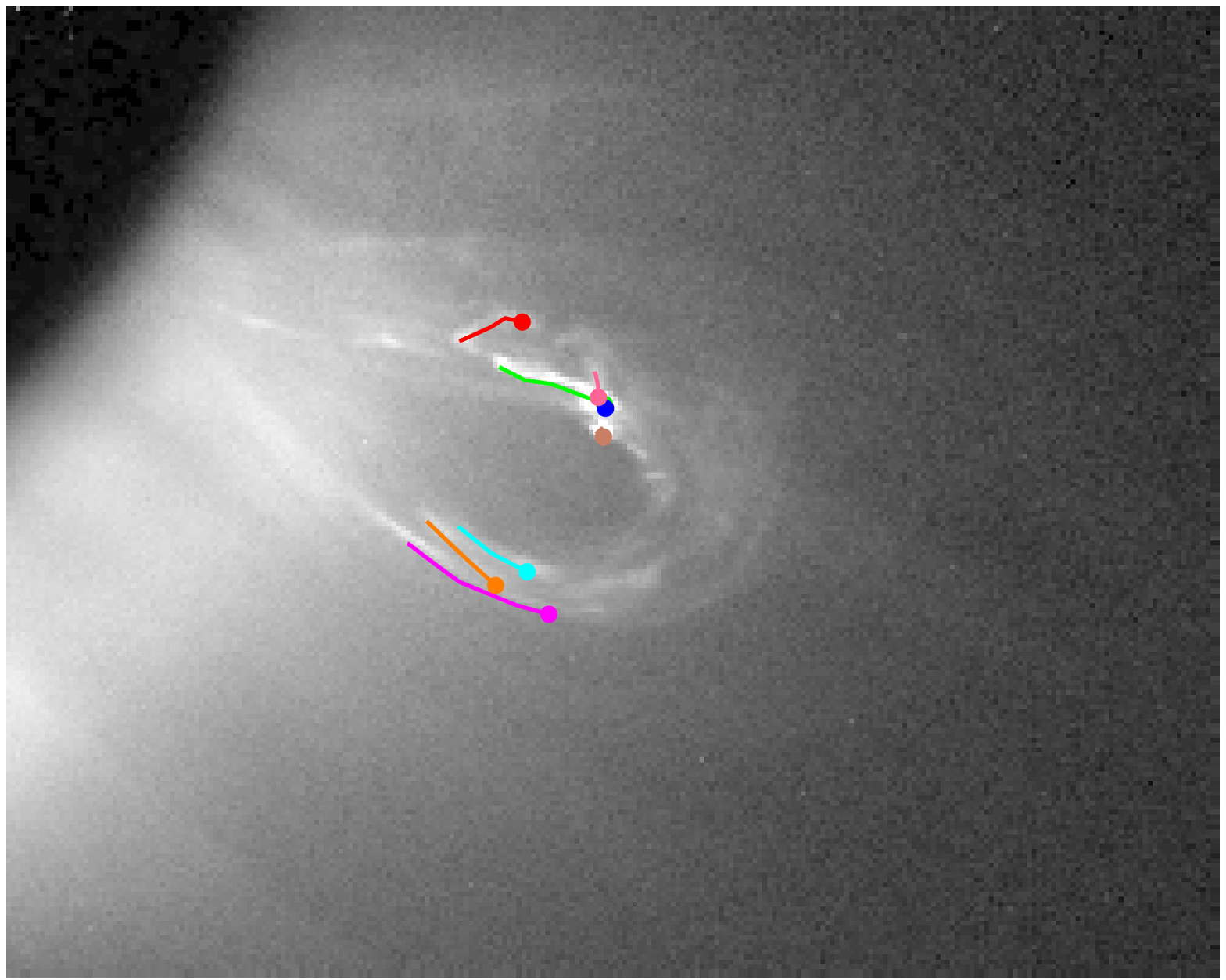}
\includegraphics[width=0.4\textwidth]{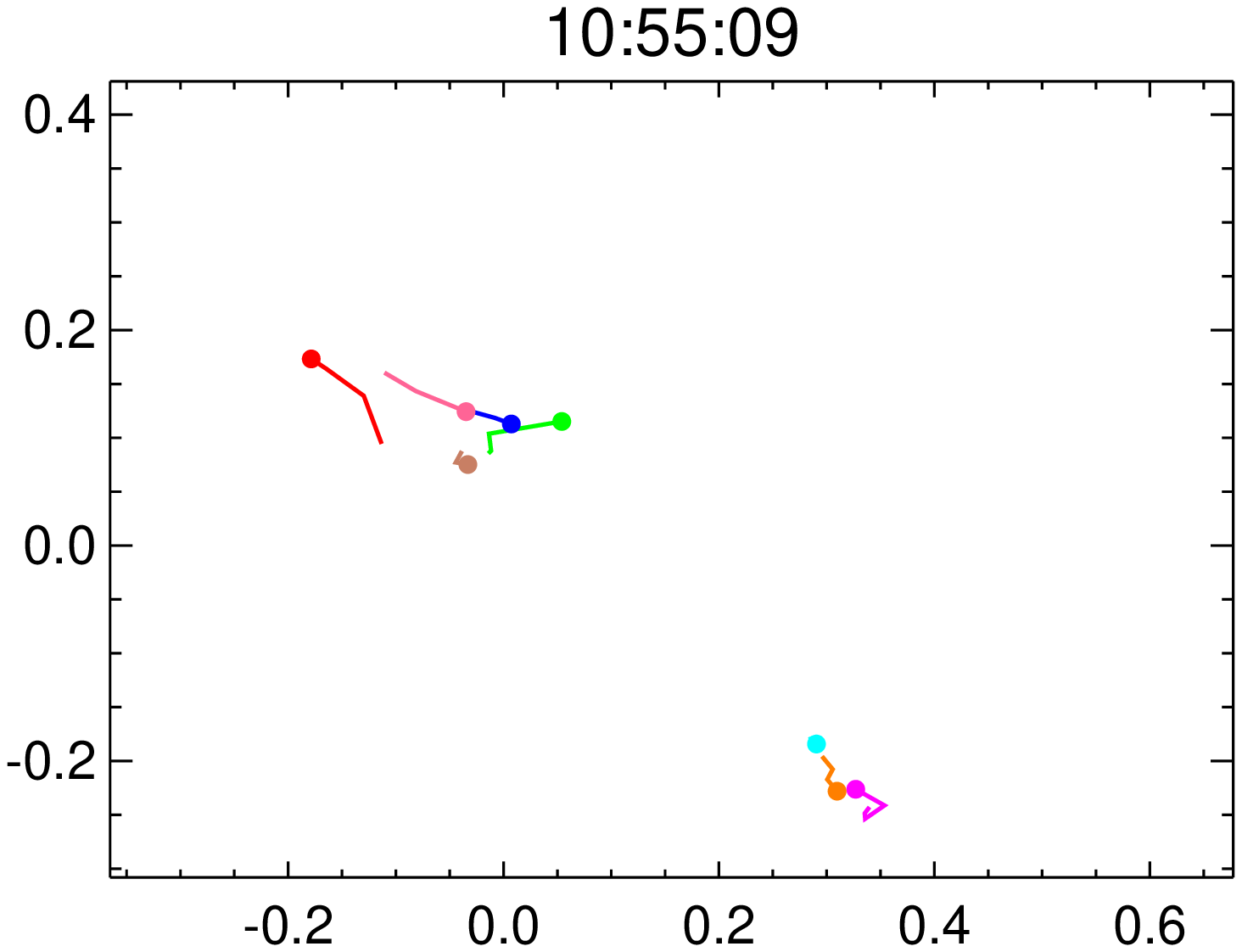}\\
\includegraphics[width=0.4\textwidth]{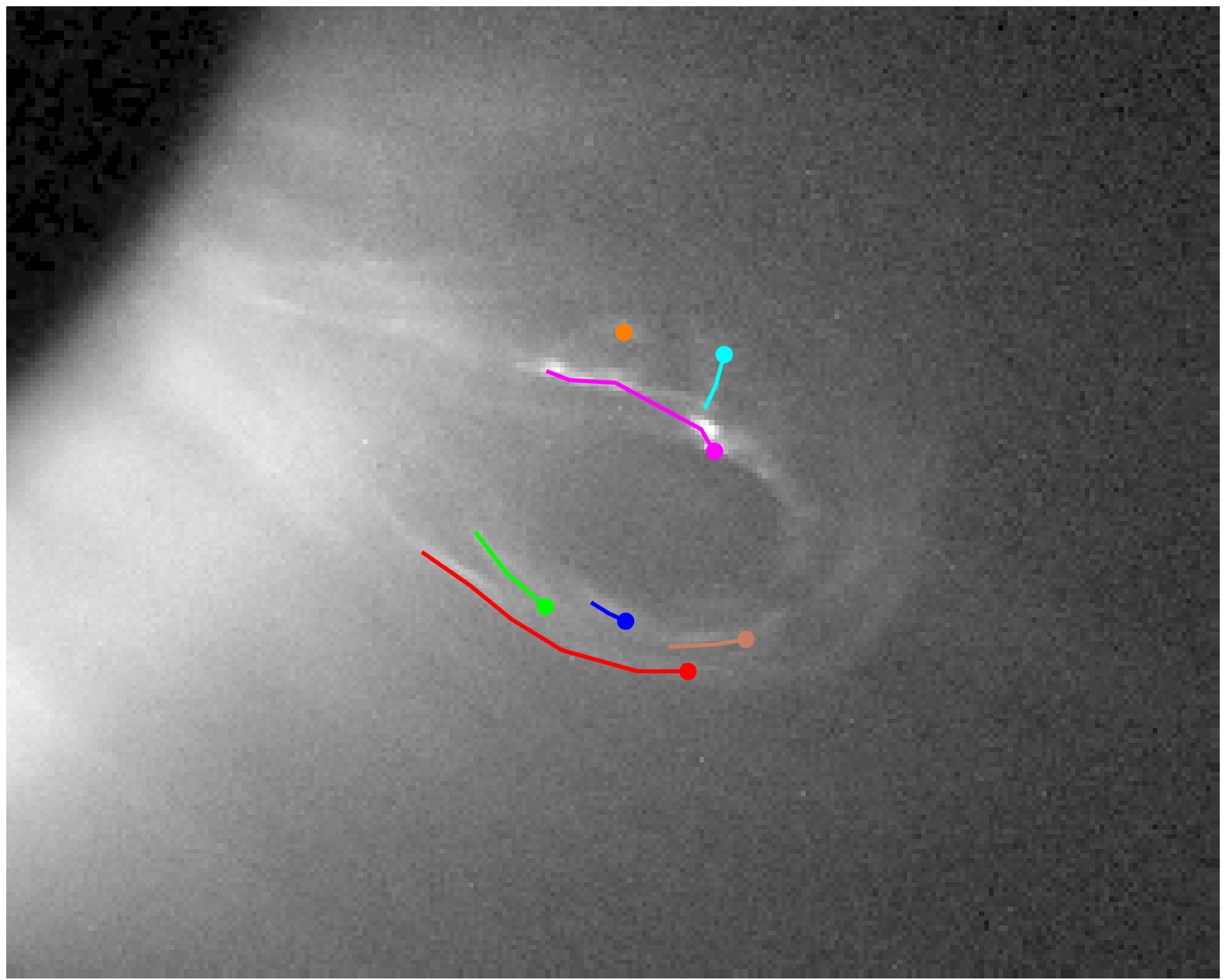}
\includegraphics[width=0.4\textwidth]{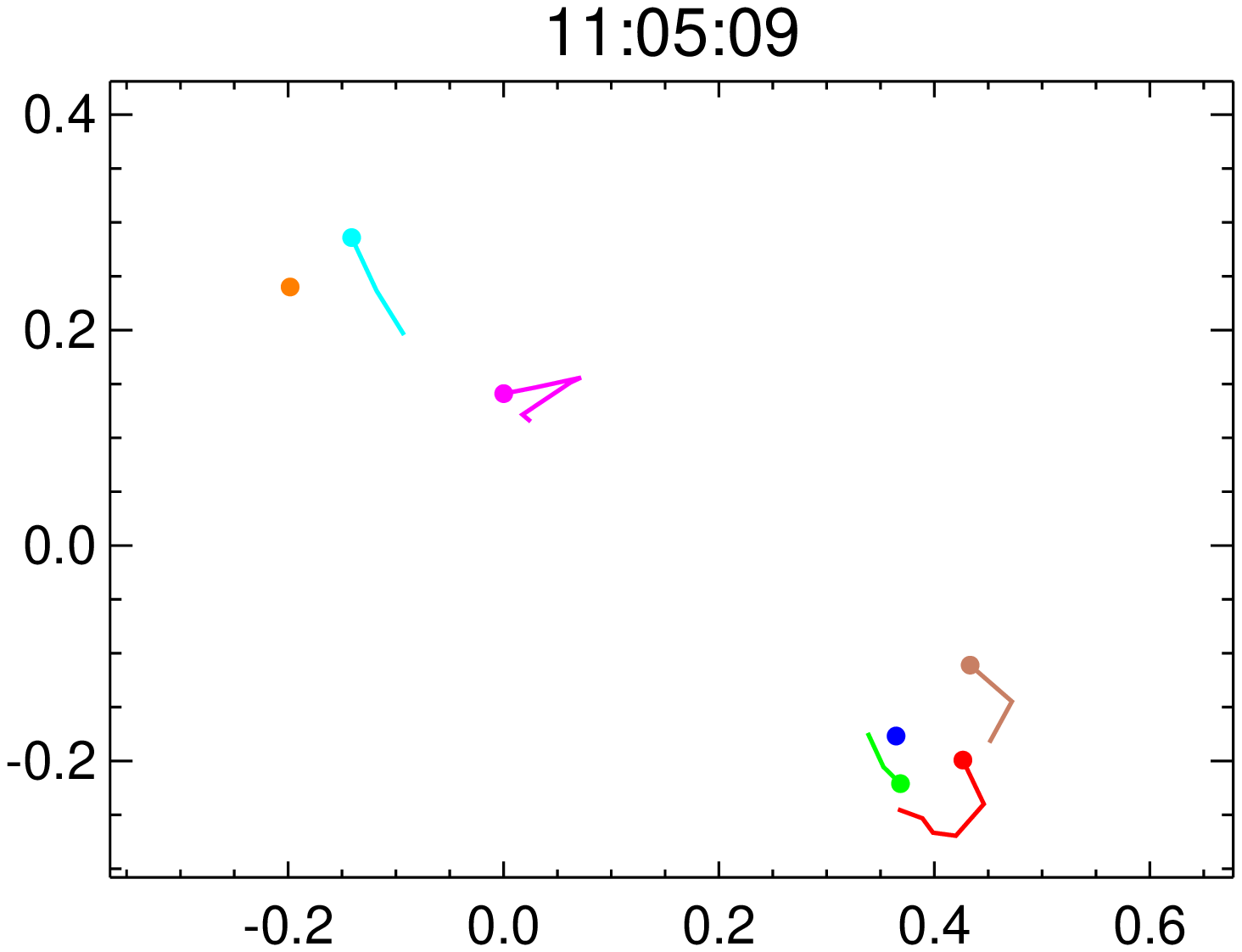}\\
\includegraphics[width=0.4\textwidth]{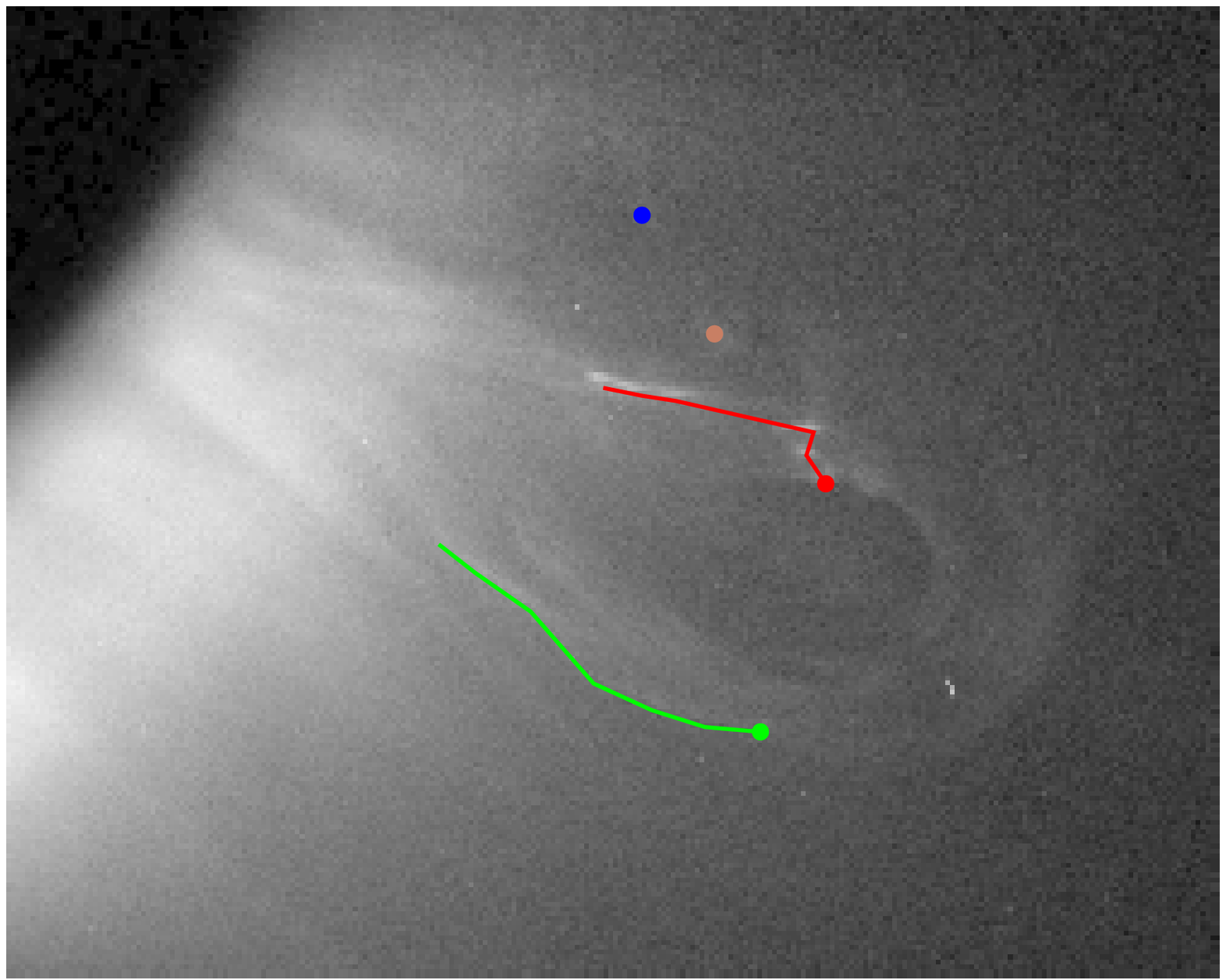}
\includegraphics[width=0.4\textwidth]{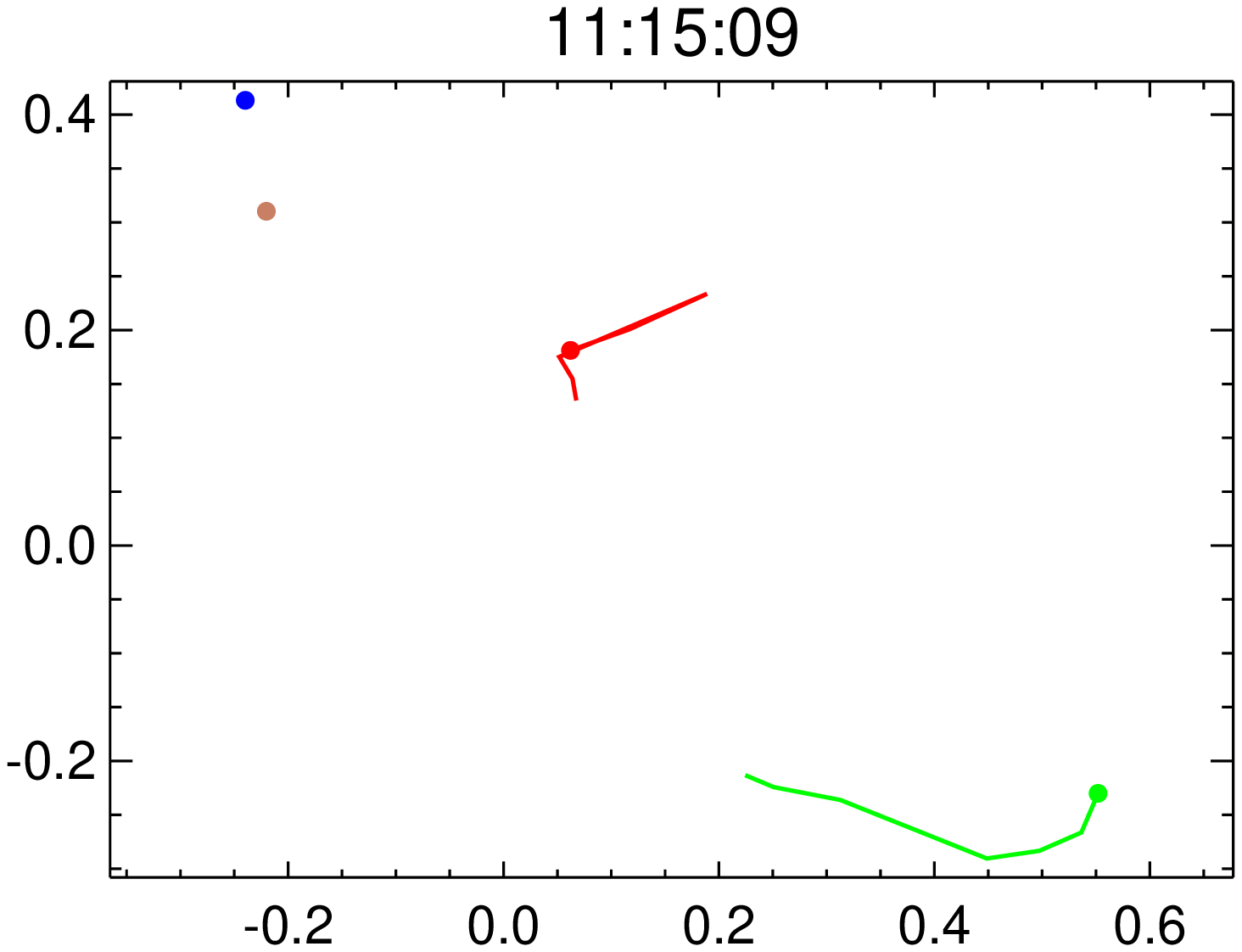}\\
\includegraphics[width=0.4\textwidth]{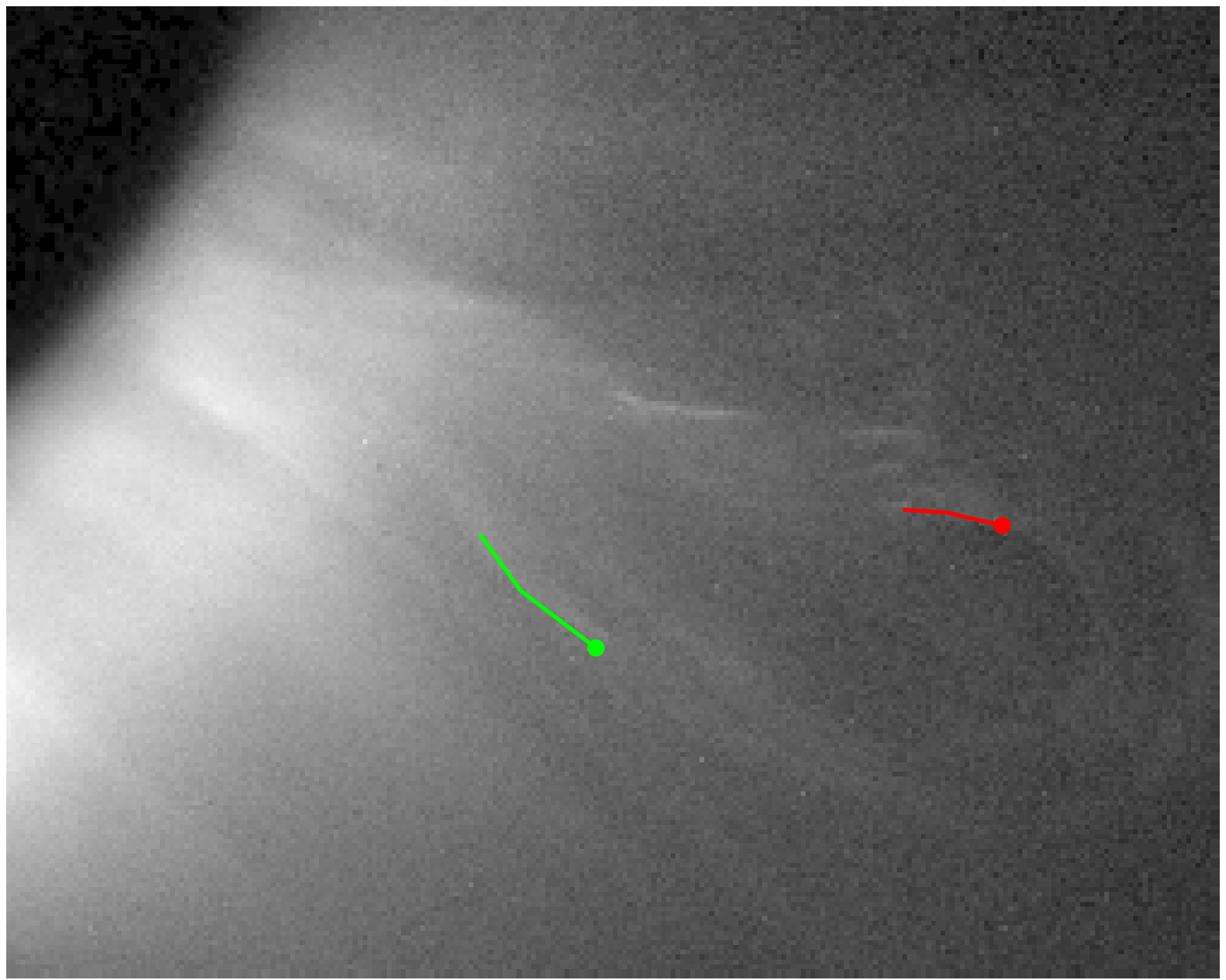}
\includegraphics[width=0.4\textwidth]{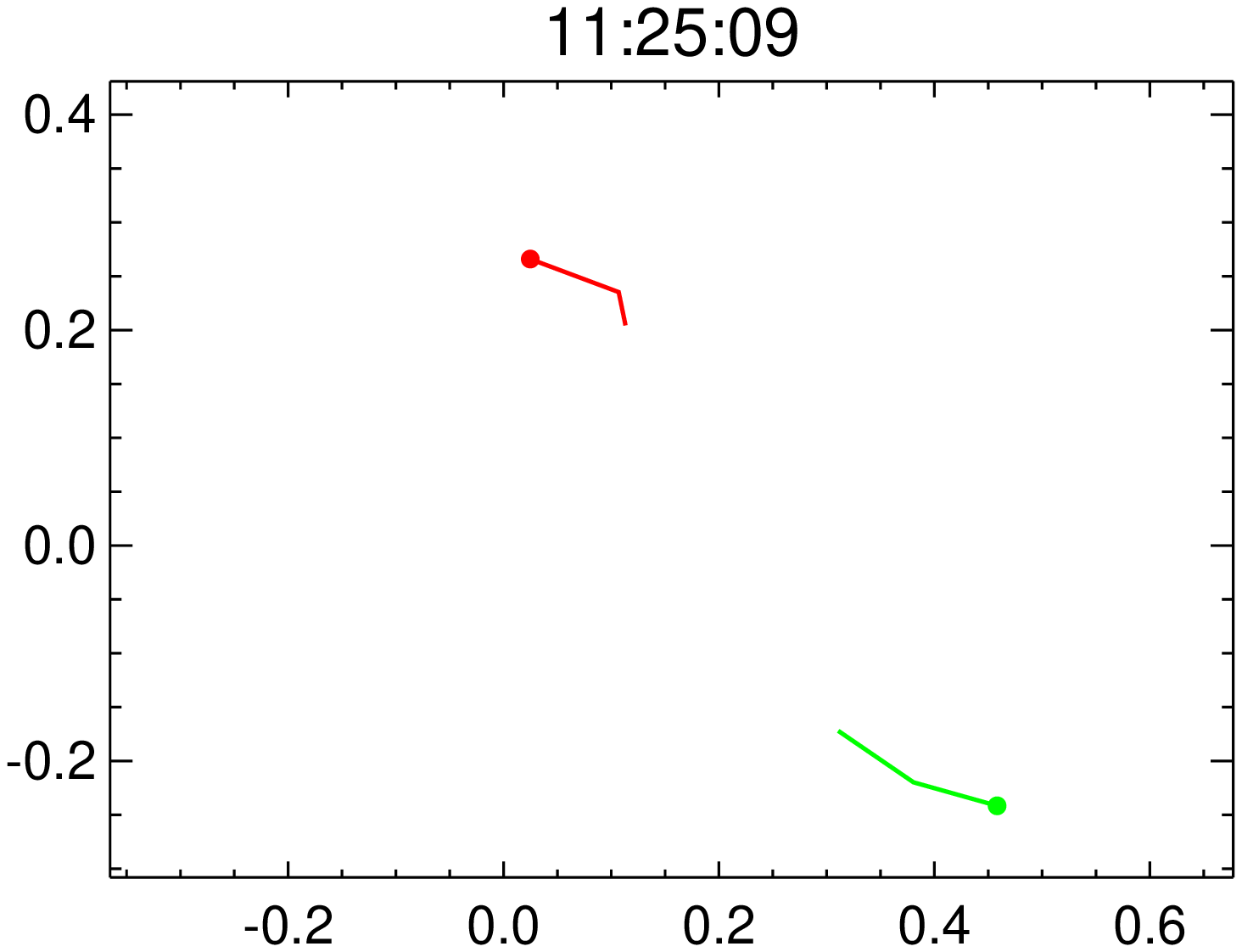}\\
\end{center}
\caption[]
{Images and plots of the prominence eruption from 10:45 to 11:25 UT, as seen in
the COR1 telescopes.  Along the left are shown the COR1-{\it Ahead} images with the
tracked features overplotted.  The plots along the right show the same data
reprojected to a viewpoint at Stonyhurst longitude 98{\degrees} west (relative
to Earth) and latitude 24{\degrees} south.  Axes are in units of solar
radii.}
\label{cor1}
\end{figure}

The EUV observations of the early part of the eruption also show a shift of the
entire prominence structure southward, from about -14{\degrees} to
-24{\degrees} latitude. Compared with the earlier
H{$\alpha$} observations (Figure~\ref{filament}), the prominence must have also
shifted $\sim15\degrees$ eastward in the early phase of the eruption.
These initial motions were followed by
radial propagation at the new latitude and longitude, $\approx$\,98W24S.
This is consistent with the analysis of \citeauthor{savage:2010} (2010)
and \inlinecite{Patsourakos&Vourlidas2011}. The latter authors fitted a
croissant-shaped flux rope model to COR2 images from both STEREO satellites
and found that the radial propagation of the CME projects back to a position
$11\degrees\pm5\degrees$ behind the limb, as seen from Earth, and
$17\degrees\pm3\degrees$ south. \inlinecite{savage:2010}
quote a heliographic position of 113{\degrees} for the associated active region
at the time of the eruption, and describe the initial motion of the prominence
as being toward the observer. 
The measurements shown in Figures~\ref{euvi1} and~\ref{euvi2} were derived
from the radial phase of the eruption.

The same analysis performed on EUVI was also performed on the COR1 images.
{At the location of the eruption, the COR1-{\it Ahead} occulter edge is at 1.37 solar
radii, and the COR1-{\it Behind} edge is at 1.59 solar radii.  The features seen by COR1
are a continuation  to higher radial distances of the truncated 304~{\AA}
features seen by EUVI.  There is a small region of overlap between EUVI and
COR1 on both spacecraft, and the observations agree within this region.  The COR1}
results are shown in Figure~\ref{cor1}.  The prominence is seen to maintain the
orientation seen in the final EUVI images. There is a slight indication of a
shift back in the clockwise direction. However, the orientations for the
final two time steps at 11:15 and 11:25~UT are based on only parts of the
prominence legs, so the apparent reverse rotation may not truly represent
the behavior of the loop as a whole.

\inlinecite{Patsourakos&Vourlidas2011} estimated the orientation of
the CME flux rope from their croissant model fit at a leading-edge
heliocentric distance of about $13~R_\odot$. 
They found that the flux rope (croissant) axis was inclined to the
east-west direction on the Sun by a tilt angle of
$-4\degrees\pm7\degrees$. This indicates a further
counter-clockwise rotation to a total value of $150\degrees\pm7\degrees$
from the original orientation estimated in Section~\ref{source_region}.
(The other possible interpretation of a clockwise rotation by
$\approx-145\degrees$ following the counter-clockwise rotation in the EUVI
height range appears far less likely.) It should be noted that the
croissant model does not include any writhe of its axis, so that the
uncertainties may be higher; however, this is not expected to change the
result by a large amount. The rotation angle at large heights represents
a close alignment of the CME flux rope with the heliospheric current
sheet (Figure~\ref{pfss}), which corresponds to the suggestion in
\inlinecite{Yurchyshyn2008} and \inlinecite{Yurchyshyn&al2009}. 

To our knowledge, the rotation of the prominence by
$\approx115\degrees$ in the corona and by $\approx150\degrees$
up to $13~R_\odot$ belongs to the largest ever inferred in this
height range. \inlinecite{Yurchyshyn&al2009} report one case of rotation
by $\approx143\degrees$  and six cases in the range
80{\degrees}--100{\degrees} out of a sample of 101 halo CMEs observed
by SOHO/LASCO, \ie, up to a distance of $30~R_\odot$.
The largest rotation found at 1~AU is on the order of $160\degrees$
\cite{Dasso&al2007,Harra&al2007}.

It is also of interest to note that the COR1 images resolve the basic
structure of the dense prominence material in the core of the CME
out to a heliocentric distance of four solar radii. The CME core has the
structure of a weakly to moderately twisted flux rope: a single flux loop
with threads that are systematically but only weakly to moderately
inclined to the axis of the loop. These indications of twist prove
to be the primary observational finding that sets a preference for a
slightly kink-unstable flux rope above a kink-stable flux rope, which
match the observed rise and rotation characteristics to a comparable
degree in the numerical modeling of the event in Paper~II.
Only few observations have revealed a flux rope structure
for an erupting prominence so clearly in this height range (see, \eg,
\opencite{Plunkett&al2000}). The images also
indicate a nearly self-similar evolution of the flux
rope throughout the instrument's field of view of (1.5--4)~$R_\odot$. The
outer coronagraph COR2 imaged the CME core as well but did not resolve
the details of its structure.

\subsection{Rotation and Rise Profile}
\label{obs_profiles}

\begin{figure}
\begin{center}
\includegraphics[width=0.9\textwidth]{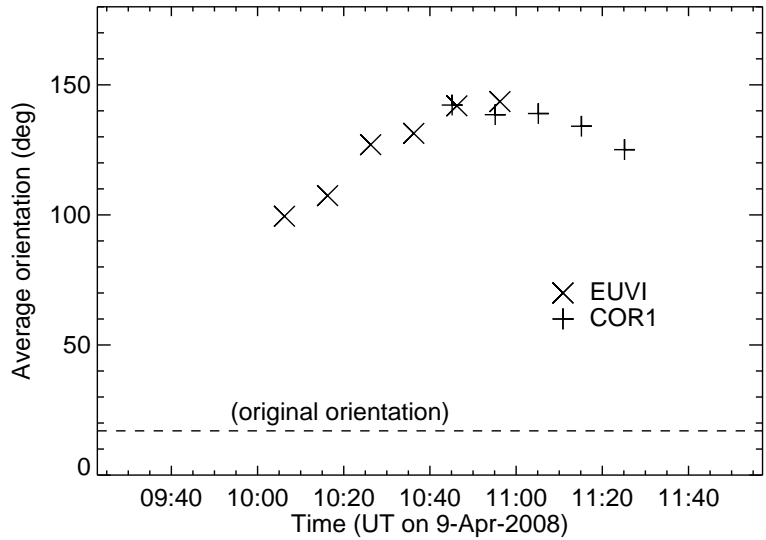}
\end{center}
\caption[]
{Orientation of the prominence body as a function of time.
The dashed line represents the original orientation as estimated from
the EUVI-{\it Ahead} 171~{\AA} image on 6~April in Figure~\ref{reformed_filament}.
The points at 10:16 and 10:26~UT are considered to be less reliable than
the others---see text.}
\label{orientation}
\end{figure}

Figure~\ref{orientation} summarizes the rotation of the erupting
prominence as a function of time from both EUVI and COR1.  The values plotted
were obtained by averaging the orientations of the individual threads whose 3D
positions could be reconstructed. The values obtained at 10:16 and
10:26~UT are considered to be less reliable than the others
in the figure, since at these times the
threads which could be located in 3D do not appear to be organized in
a well defined structure, as, for example, flux rope legs (see the
right panels in Figure~\ref{euvi1}).
{It was not possible to derive the 3D prominence orientation for times before
10:06~UT due to source confusion.  However, the visible appearance of the
prominence as seen from STEREO {\it Ahead} between 9-10 UT is consistent with
considerable rotation throughout this time period, leading to the conclusion that
most of the rotation occured during the initial phase of the
eruption.}

\begin{figure}
\begin{center}
\includegraphics[width=0.9\textwidth]{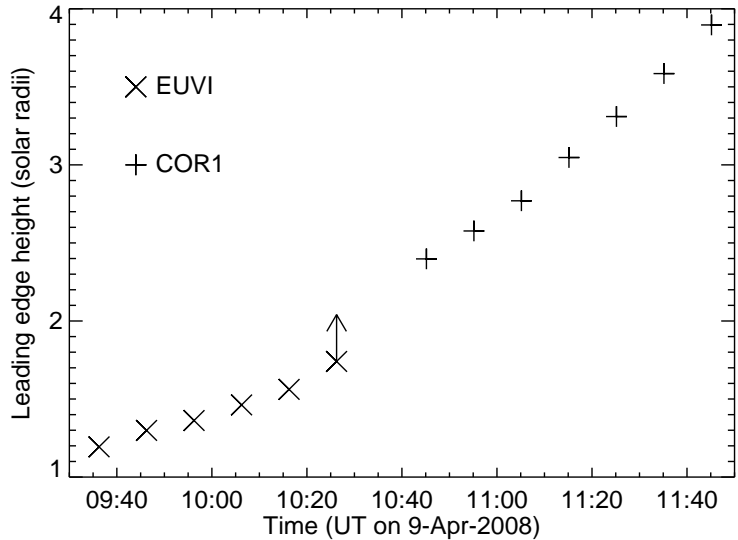}
\end{center}
\caption[]
{Heliocentric heights of the prominence leading edge \vs\ time. The value
for 10:26~UT is shown as a lower limit, since the prominence is right at the
edge of the EUVI-{\it Ahead} field-of-view at this time.}
\label{looptop}
\end{figure}

We also used \verb+scc_measure+ to derive a time-height curve for the
prominence eruption.  This was done by matching points at the leading edge of
the prominence in both images.  Since these points in the two images are not
guaranteed to represent exactly the same point in 3D space, the results can
only be treated as an approximation.  However, since the extent of the
prominence material is relatively small, the approximation is quite valid.  The
resulting time-height curve for the top of the prominence is shown in
Figure~\ref{looptop}.  The COR1 values appear to be a bit higher than one would
expect from an extrapolation of the EUVI data.  That may represent a difference
between the appearance of the prominence at 304~{\AA} versus white light.  
In addition, the heights derived from EUVI
data after 9:56~UT may systematically fall short of the true heights.
{It is possible that the traceable threads did not
extend up to the true top of the structure in these images, since
an upward extension of the loop-shaped structure in the image at
9:56~UT had faded by 10:06~UT.} The estimated height at 10:26~UT {is
only a lower limit given} by the edge of the EUVI-{\it Ahead} field-of-view.

A quadratic fit to the EUVI data gives an initial speed of \mbox{67 km
s$^{-1}$}, accelerating to \mbox{173 km s$^{-1}$} as the prominence leaves
the EUVI
field-of-view.  A quadratic fit to the heights measured by COR1 gives a
velocity starting at \mbox{206 km s$^{-1}$}, and accelerating to \mbox{379 km
s$^{-1}$} by the end of the COR1 data.  There's some evidence from the
COR1 quadratic fit that the acceleration strongly decreased when the leading
edge of the prominence reached a heliocentric height of 3~solar radii.  
This is borne out by a linear fit to the COR1 data points above 3~solar radii,
which gives a slightly lower velocity of \mbox{327 $\pm$ 9.2 km s$^{-1}$}.
The combined EUVI and COR1 data in Figure~\ref{looptop} give the visual
impression that much of the acceleration actually occurred up to
$\approx\!2.5$~solar radii.
The COR2 data give a speed close to \mbox{400 km s$^{-1}$}.
All of these values refer to estimated positions of the apex point of
the visible flux rope structure in the core of the CME.

\citeauthor{Landi&al2010} (2010) also derive a time-height profile for the
leading edge of the prominence material, using a combination of STEREO
{\it Ahead} and SOHO data.  Their results are very close to those presented
here, with the possible exception of the amount of acceleration.
\citeauthor{Landi&al2010} quote an acceleration of \mbox{59.6 m s$^{-2}$},
while our results are more consistent with a smaller value of
\mbox{37.9 $\pm$ 4.0 m s$^{-2}$} over the time range covered by
Figure~\ref{looptop}. They also give the projected velocity of
the CME leading edge in the plane of the sky of STEREO {\it Ahead},
which approaches \mbox{700 km s$^{-1}$}, still rising, at a heliocentric
height of $3~R_\odot$.

\begin{figure}
\begin{center}
\includegraphics[width=0.9\textwidth]{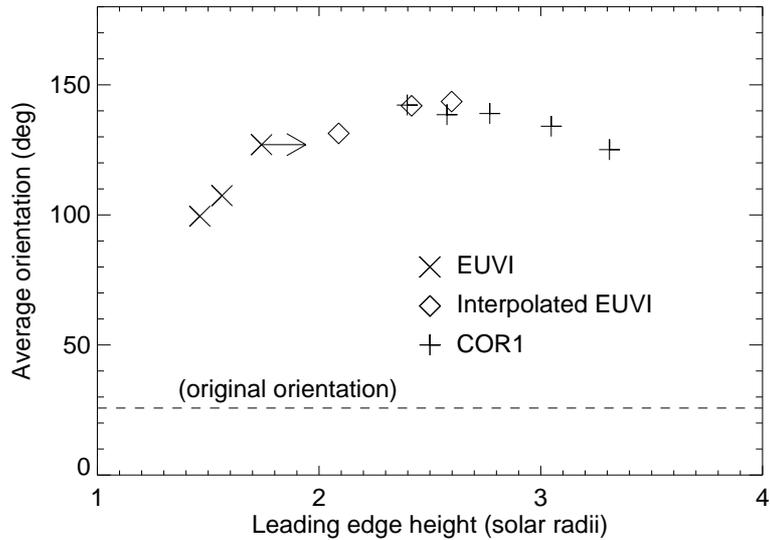}
\end{center}
\caption[]
{Prominence rotation \vs\ heliocentric height of its leading edge.  The
diamond symbols represent EUVI and COR1 height data from Figure~\ref{looptop}
interpolated to the times of EUVI orientation measurements in
Figure~\ref{orientation}.}
\label{rotation}
\end{figure}

Combining the information in Figures~\ref{orientation} and
\ref{looptop}, we obtain the rotation of the prominence as a function of
its leading edge height, see Figure~\ref{rotation}. (As there are only
three time
steps in common between the EUVI data sets, we have interpolated the EUVI and COR1
height data from Figure~\ref{looptop} to the times of the EUVI measurements in
Figure~\ref{orientation} to better show the trend.) These data emphasize again
that much of the rotation is acquired in the height range
$h\lesssim R_\odot/2$, or $\sim\!2$ initial footpoint separations,
above the photosphere.

{
\subsection{Comparison With the Orientation of the PIL}
\label{PIL_orientation}

The comparison of the PFSS extrapolation with the orientation of the
prominence at large distances from the Sun obtained in
\inlinecite{Patsourakos&Vourlidas2011} shows that the prominence has
approached a close alignment with the heliospheric current sheet
(Section~\ref{ss:measurements}). An obvious question is whether such
alignment could also lead to the rotation in the coronal height range
studied here. We will see in Paper~II that the rotation of erupting
flux ropes and the change of the PIL orientation with height always
point in opposite directions if the eruption originates in a simple
bipolar region. However, Figure~\ref{pfss2} shows that the topology of
the field above AR~10989 is more complex. In particular, the field
component along the filament channel, which influences the rotation of
erupting flux \cite{isenberg:2007}, changes sign with increasing
height. In Figure~\ref{pfss2}, it is
clearly seen that the relevant section of the PIL keeps a nearly
constant orientation up to about 1.15~$R_\odot$, above which it starts
to shrink noticeably, disappearing completely by 1.3~$R_\odot$.
Thus, any rotation in the height range up to about 1.15~$R_\odot$
could not be caused by a changing PIL direction. In the height range
$\approx(1.15\mbox{--}1.3)~R_\odot$ the prominence did not see a PIL
with a well defined direction. Above 1.3~$R_\odot$, the prominence
entered the large-scale field structure defined by the polar fields
and the heliospheric current sheet. The PIL direction in this range is
relatively similar to
the direction of the
photospheric PIL in the active region and the
original orientation of the prominence. Hence, throughout the height
range of strong initial prominence rotation,
$\sim(1\mbox{--}1.5)~R_\odot$, the assumption of rotation by alignment
with the relevant PIL is inconsistent with the structure of the field.
The prominence actually rotated away from the PIL in this height
range. 

The situation changed when the initial prominence rotation began to
level off at a large angle. The subsequent rotation up to the height
considered in \inlinecite{Patsourakos&Vourlidas2011} is consistent with
an alignment with the PIL in the heliospheric current sheet. From the
orientation reached in the initial phase, it was favorable for the
prominence to continue the rotation to align with the PIL nearly
antiparallel to the original orientation.

These conclusions are robust against the limitations of
the PFSS model, which reflects the magnetic structure in and around
AR~10989 primarily at the time of central meridian passage, since the
region did not show any strong magnetic changes in the time before
the eruption on 9~April. The orientation of the PIL in the
photosphere, as outlined by the prominence in the EUV, changed by
about 20{\degrees}, much smaller than the observed prominence rotation
in the low corona. No signs were seen of any major flux emergence
which might have altered the topology of the ring-shaped PIL seen
at low coronal heights.

Generally, one expects a dominance of the Lorentz force in driving any
rotation of erupting flux at low coronal heights, where the plasma
beta is small. As summarized in the following section and detailed in
Paper~II, the Lorentz force due to both the tension of twisted field
lines and the presence of a shear field component in the ambient field
causes a rotation away from the PIL, along which the rising flux is
originally oriented. As larger heights are reached, the currents and
Lorentz forces in the erupting flux decrease, while the influence of
the pressure gradient force increases with the increasing plasma beta
in the ambient field. The alignment with the heliospheric current
sheet then reaches an increasing importance in the dynamics of the
eruption. The transition is expected to occur when the plasma beta
approaches unity.

A difference between the PIL in AR~10989 and the large-scale PIL in
the streamer should be noted. While the PILs are nearly parallel to
each other, the field direction across them is reversed
(Figure~\ref{pfss2}). The horizontal field component across the PIL in
the streamer exerted a Lorentz force on the prominence legs according to
the mechanism proposed in \inlinecite{isenberg:2007}, because the
prominence was directed approximately along this field
component (perpendicular to the PIL) after the initial rotation had
leveled off, \ie, in the whole COR1 height range.
With the current flowing opposite to the axial field of the dextral
prominence, the direction of this force on the legs of the prominence is
such that the top part rotates in the clockwise direction. This might
explain the slight backward rotation indicated by the COR1 data.
If this were the dominant effect in the subsequent evolution,
the initial rotation would be reversed until the prominence
aligns with the heliospheric current sheet in the direction parallel
to its original orientation. Since the transition to high-beta
conditions occurs in the COR1-COR2 height range, it is
unlikely that the Lorentz force could dominate the subsequent
rotation. Consequently, antiparallel alignment with the PIL in the
heliospheric current sheet due to the pressure gradient force, which
drives such alignment on the shortest possible path, independent of
the field direction, appears to be the most likely evolution in the
COR2 height range.
}

\section{Numerical Modeling}

\begin{figure}
\begin{center}
\includegraphics[width=0.63\textwidth]{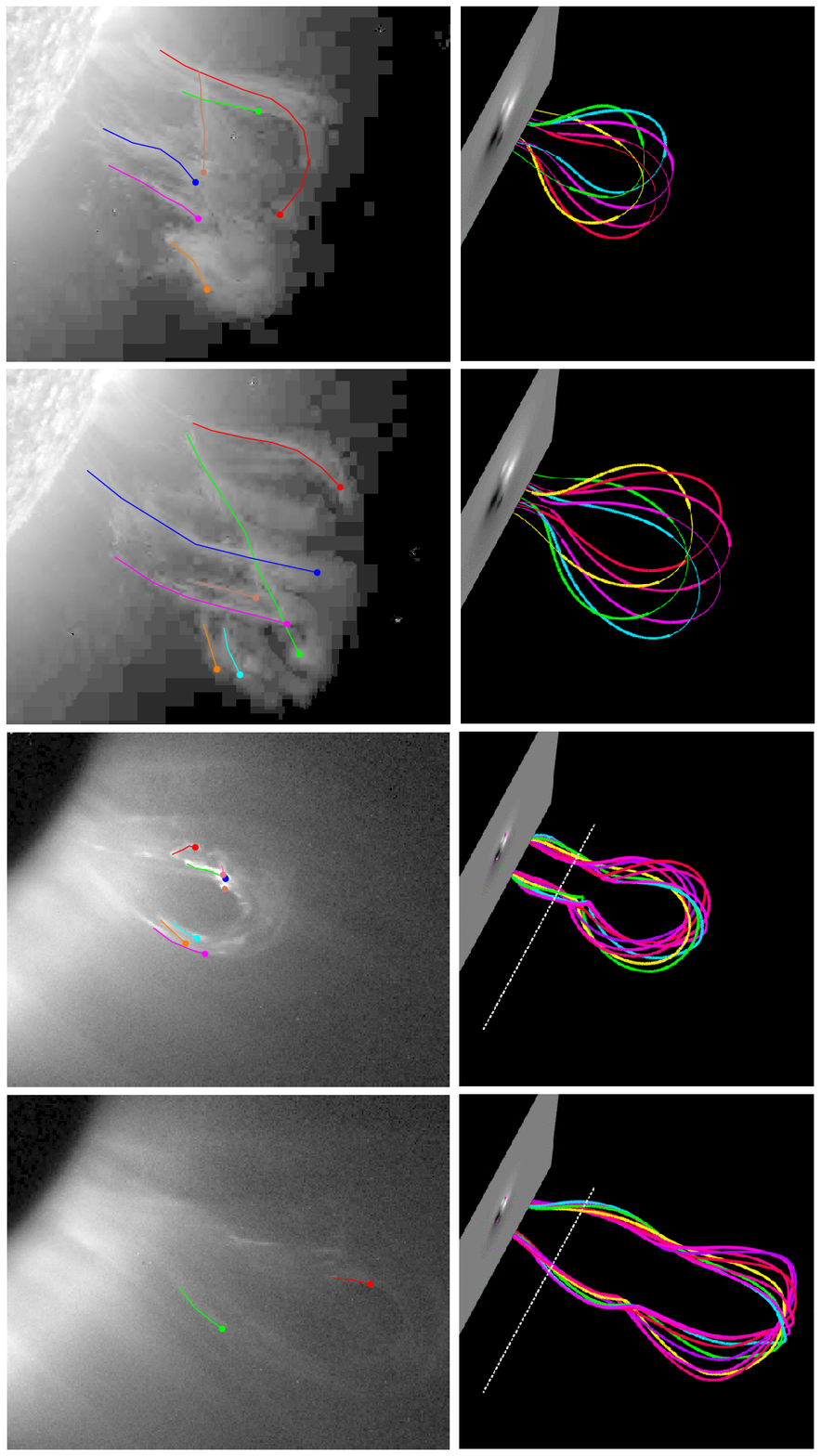}
\end{center}
\caption[]
{Shape of the erupting flux in the best-matching numerical model --- a
 weakly kink-unstable flux rope of initial twist $3.5\pi$ --- at various
 stages of the rise, compared with STEREO {\it Ahead} images at
 10:16, 10:26, 10:55, and 11:25~UT.
 A set of field lines enclosing the magnetic axis of the simulated
 flux rope is plotted at the times the rope apex has reached the
 corresponding heights, $h=7.3$, 9.6, 20, and $30~h_0$ ($h_0$ -- initial
 apex height), which are obtained from the scaling of the footpoint
 distance, $3.3~h_0$, to the value of 175~Mm estimated in
 Section~\ref{source_region}. The observed and simulated flux ropes are
 displayed at the same perspective: the line connecting the footpoints
 of the rope makes an angle of 26{\degrees} with the line of sight and
 the vertical axis in the simulation is tilted away from the observer
 by 8{\degrees} (so that the magnetogram, $B_z(x,y,0,t)$, is seen from
 the bottom side). The dotted line indicates the edge of the COR1
 occulting disk.
 Since the visible threads in the EUVI images do not outline the
 complete shape of the flux rope, sections of the field lines in the
 corresponding simulation snapshots are highlighted for better visual
 correspondence. For each field line, 200 segments at the top and 150
 segments at each bottom end are plotted with reduced line width. The
 segments are obtained from a numerical integration along the field
 line with adaptive step size and are all different, resulting in
 different lengths of the highlighted sections, analogous to the
 observations.}
\label{compare}
\end{figure}

Paper~II presents a parametric study of various effects which lead to,
or influence, the rotation of ejected flux ropes {in low-beta plasma}.
The initial flux rope twist, the strength of the external shear field
component, and the height profile of the overlying potential field
are varied. The latter is related to
the distance between the main polarities in the photosphere.
Figure~\ref{compare} shows field line plots from the
numerical model that best fits the observations in their entirety
(rise and rotation profiles and the STEREO images). In
this model, a moderate initial twist of $\Phi=3.5\pi$ triggers a weak
helical kink instability, which lifts the initially force-free,
toroidal flux rope into the torus-unstable range of heights and also
contributes to the rotation. The ejection is primarily driven by the
torus instability \cite{kliem:2006}, so that the model belongs to the
loss-of-equilibrium category \cite{Priest&Forbes2002}. The major part
of the rotation in this run is due to the presence of an external
shear field component (pointing along the prominence and polarity
inversion line and due to sources outside the flux rope), as first
suggested in \inlinecite{isenberg:2007}. The strength of the shear
field at the initial flux rope apex is $2/3$ of the external
poloidal field which holds the rope in equilibrium.
{The value is consistent with a very rough estimate of this ratio
from the structure of the active region. The ratio of the distances
between the main polarities along and across the PIL is about 1/2 in
the PFSS field, and the shear should slightly increase with the
gradual change of the PIL direction in the time till the eruption.}

The overall shapes of the flux rope in the considerable height range
included in Figure~\ref{compare}, as well as the inclination of the
still weakly twisted field lines to the axis of the rope at the
largest height, match the observations reasonably well. (Note that the
rope has acquired a large part of its total rotation already at the
first time selected; see Figure~\ref{orientation}.) Similar agreement
is demonstrated in Paper~II for the height-rotation and time-height
profiles.

A model with subcritical initial twist, $\Phi=2.5\pi$, which allows
only for the development of the torus instability, reproduces the rise
profile similarly well, while the match with the rotation profile is
somewhat worse. The field lines at the strongly expanded stage of the
final COR1 image in Figure~\ref{compare} are rather straight and
hardly show any indication of the observed twist. Moreover, this flux
rope requires a considerable initial perturbation to reach the
torus-unstable range of heights, which is not supported by the
pre-eruption height estimate and the initial dynamics of the
prominence described in Section~\ref{data_analysis} (see Paper~II).

\section{Summary and Conclusions}

The images of the STEREO EUVI and COR1 telescopes resolve the basic
structure of the erupting prominence in the core of the
``Cartwheel CME'' on 9~April 2008 out to heliocentric distances of
$4~R_\odot$. A single flux rope with indications of weak to moderate twist
is revealed. The flux rope expands approximately self-similarly in the
COR1 field of view of (1.5--4)~$R_\odot$.

The true path of the prominence could be reconstructed from
stereoscopic observations even when the
angular separation between the STEREO spacecraft was as much as
48{\degrees}.  The reconstruction reveals that the {dextral}
prominence rotated counter-clockwise
by a large angle of $\approx\!115\degrees$ up to a heliocentric height
of $2.5~R_\odot$, where the rotation leveled off. A gentle backward
rotation by $\approx\!15${\degrees} may have followed in the height
range up to $3.3~R_\odot$. Two thirds of the rotation
were acquired within $0.5~R_\odot$ from the photosphere. The coronal
height profile of the rotation angle in a CME
is thus derived for the first time. To our knowledge, this
ranks as one of
the largest rotations so far measured in
the corona.

Taken jointly with recent results about the orientation of the CME at a
heliocentric distance of $\approx\!13~R_\odot$, obtained from fitting a
croissant-shaped flux rope model to stereoscopic images from COR2
\cite{Patsourakos&Vourlidas2011}, a further counter-clockwise rotation
by $\approx\!35\degrees$ from the value at $2.5~R_\odot$ is indicated.
This aligned the erupted flux closely with the heliospheric
current sheet above the active region.
{However, the structure of the field within the first
$\sim\!0.5~R_\odot$ above the original location of the prominence,
obtained from a PFSS extrapolation, precludes alignment with the PIL
as mechanism for the initial rotation.}


The rotation and rise profiles of the prominence in the EUVI-COR1
height range are reproduced by a numerical model that follows the
evolution of an unstable force-free flux rope. A parametric study of
flux rope rotation in this model, detailed in Paper~II, suggests that the main
part of the rotation in
{this height range} was caused by the shear field in the source
volume. While the rotation and rise profiles can be modeled nearly
equally well by a weakly kink-unstable flux rope of $3.5\pi$ initial
twist and by a kink-stable flux rope of $2.5\pi$ initial twist, the
indications of twist in the COR1 images and the absence of a strong
initial perturbation in the EUVI-{\it Ahead} data favor the $3.5\pi$ model,
suggesting that the helical kink instability did occur and contributed
the remaining part of the {rotation}.

{In summary, the large total rotation of the erupting flux rope in the
Cartwheel CME is due to a combination of three effects. The internal
Lorentz forces associated with the shear field and with the tension of
twisted field lines caused a strong rotation in the inner corona. The
resulting orientation relative to the heliospheric current sheet was
favorable for the subsequent alignment with the sheet in the outer
corona and inner solar wind to proceed with the same sense of
rotation. Since the heliospheric current sheet had an orientation
relatively near the original direction of the erupting flux, a very
large final rotation angle was reached.}

\acknowledgements
We acknowlege the use of data provided by the Global High Resolution H$\alpha$
Network, and by the SECCHI instruments on the STEREO spacecraft.  
Magnetic field extrapolations were supplied by the CCMC; special thanks to
Lutz Rastaetter and Peter MacNeice for help with the CCMC data.
We thank {the anonymous referee for a constructive report which led
to a deeper consideration of PIL orientation \vs\ height, and}
S.~Patsourakos for information about his fitting of the
CME orientation in the COR2 height range.
WTT's work was supported by NASA Grant NNG06EB68C.
BK's work was supported by the DFG, the STFC, and by NASA through Grant
NNX08AG44G.
TT's work was partially supported by the European Commission
through the SOTERIA Network (EU FP7 Space Science Project No.\ 218816)
and by the NASA HTP and LWS programs.

\bibliography{prominence}

\begin{thebibliography}{30}
\ifx \bisbn   \undefined \def \bisbn  #1{ISBN #1}\fi
\ifx \binits  \undefined \def \binits#1{#1}\fi
\ifx \bauthor  \undefined \def \bauthor#1{#1}\fi
\ifx \batitle  \undefined \def \batitle#1{#1}\fi
\ifx \bjtitle  \undefined \def \bjtitle#1{\textit{#1}}\fi
\ifx \bvolume  \undefined \def \bvolume#1{\textbf{#1}}\fi
\ifx \byear  \undefined \def \byear#1{#1}\fi
\ifx \bissue  \undefined \def \bissue#1{#1}\fi
\ifx \bfpage  \undefined \def \bfpage#1{#1}\fi
\ifx \blpage  \undefined \def \blpage #1{#1}\fi
\ifx \burl  \undefined \def \burl#1{\textsf{#1}}\fi
\ifx \href  \undefined \def \href#1#2{\textsf{#2}}\fi
\ifx \doiurl  \undefined \def
  \doiurl#1{\href{http://dx.doi.org/#1}{\textsf{#1}}}\fi
\ifx \betal  \undefined \def \betal{\textit{et al.}}\fi
\ifx \binstitute  \undefined \def \binstitute#1{#1}\fi
\ifx \bctitle  \undefined \def \bctitle#1{#1}\fi
\ifx \beditor  \undefined \def \beditor#1{#1}\fi
\ifx \bpublisher  \undefined \def \bpublisher#1{#1}\fi
\ifx \bbtitle  \undefined \def \bbtitle#1{\textit{#1}}\fi
\ifx \bedition  \undefined \def \bedition#1{#1}\fi
\ifx \bseriesno  \undefined \def \bseriesno#1{\textbf{#1}}\fi
\ifx \blocation  \undefined \def \blocation#1{#1}\fi
\ifx \bsertitle  \undefined \def \bsertitle#1{\textit{#1}}\fi
\ifx \bsnm \undefined \def \bsnm#1{#1}\fi
\ifx \bsuffix \undefined \def \bsuffix#1{#1}\fi
\ifx \bparticle \undefined \def \bparticle#1{#1}\fi
\ifx \barticle \undefined \def \barticle#1{}\fi
\ifx \botherref \undefined \def \botherref#1{}\fi
\ifx \url \undefined \def \url#1{\textsf{#1}}\fi
\ifx \bchapter \undefined \def \bchapter#1{}\fi
\ifx \bbook \undefined \def \bbook#1{}\fi
\ifx \bcomment \undefined \def \bcomment#1{#1}\fi
\ifx \oauthor \undefined \def \oauthor#1{#1}\fi
\ifx \citeauthoryear \undefined \def \citeauthoryear#1{#1}\fi
\def \endbibitem {}
\ifx \bconflocation  \undefined \def \bconflocation#1{#1} \fi

\bibitem[\protect\citeauthoryear{{Bi} \textit{et~al.}}{2011}]{bi:2011}
\begin{barticle}
\bauthor{\bsnm{{Bi}}, \binits{Y.}},
\bauthor{\bsnm{{Jiang}}, \binits{Y.C.}},
\bauthor{\bsnm{{Yang}}, \binits{L.H.}},
\bauthor{\bsnm{{Zheng}}, \binits{R.S.}}:
\byear{2011},
\bjtitle{New Astronomy}
\bvolume{16},
\bfpage{276}.
doi:\doiurl{10.1016/j.newast.2010.11.009}.
\end{barticle}
\endbibitem

\bibitem[\protect\citeauthoryear{{Culhane}
  \textit{et~al.}}{2007}]{culhane:2007}
\begin{barticle}
\bauthor{\bsnm{{Culhane}}, \binits{J.L.}},
\bauthor{\bsnm{{Harra}}, \binits{L.K.}},
\bauthor{\bsnm{{James}}, \binits{A.M.}},
\bauthor{\bsnm{{Al-Janabi}}, \binits{K.}},
\bauthor{\bsnm{{Bradley}}, \binits{L.J.}},
\bauthor{\bsnm{{Chaudry}}, \binits{R.A.}},
\bauthor{\bsnm{{Rees}}, \binits{K.}},
\bauthor{\bsnm{{Tandy}}, \binits{J.A.}},
\bauthor{\bsnm{{Thomas}}, \binits{P.}},
\bauthor{\bsnm{{Whillock}}, \binits{M.C.R.}},
\bauthor{\bsnm{{Winter}}, \binits{B.}},
\bauthor{\bsnm{{Doschek}}, \binits{G.A.}},
\bauthor{\bsnm{{Korendyke}}, \binits{C.M.}},
\bauthor{\bsnm{{Brown}}, \binits{C.M.}},
\bauthor{\bsnm{{Myers}}, \binits{S.}},
\bauthor{\bsnm{{Mariska}}, \binits{J.}},
\bauthor{\bsnm{{Seely}}, \binits{J.}},
\bauthor{\bsnm{{Lang}}, \binits{J.}},
\bauthor{\bsnm{{Kent}}, \binits{B.J.}},
\bauthor{\bsnm{{Shaughnessy}}, \binits{B.M.}},
\bauthor{\bsnm{{Young}}, \binits{P.R.}},
\bauthor{\bsnm{{Simnett}}, \binits{G.M.}},
\bauthor{\bsnm{{Castelli}}, \binits{C.M.}},
\bauthor{\bsnm{{Mahmoud}}, \binits{S.}},
\bauthor{\bsnm{{Mapson-Menard}}, \binits{H.}},
\bauthor{\bsnm{{Probyn}}, \binits{B.J.}},
\bauthor{\bsnm{{Thomas}}, \binits{R.J.}},
\bauthor{\bsnm{{Davila}}, \binits{J.}},
\bauthor{\bsnm{{Dere}}, \binits{K.}},
\bauthor{\bsnm{{Windt}}, \binits{D.}},
\bauthor{\bsnm{{Shea}}, \binits{J.}},
\bauthor{\bsnm{{Hagood}}, \binits{R.}},
\bauthor{\bsnm{{Moye}}, \binits{R.}},
\bauthor{\bsnm{{Hara}}, \binits{H.}},
\bauthor{\bsnm{{Watanabe}}, \binits{T.}},
\bauthor{\bsnm{{Matsuzaki}}, \binits{K.}},
\bauthor{\bsnm{{Kosugi}}, \binits{T.}},
\bauthor{\bsnm{{Hansteen}}, \binits{V.}},
\bauthor{\bsnm{{Wikstol}}, \binits{{\O}.}}:
\byear{2007},
\bjtitle{Solar Phys.}
\bvolume{243},
\bfpage{19}.
doi:\doiurl{10.1007/s01007-007-0293-1}.
\end{barticle}
\endbibitem

\bibitem[\protect\citeauthoryear{{Dasso} \textit{et~al.}}{2007}]{Dasso&al2007}
\begin{barticle}
\bauthor{\bsnm{{Dasso}}, \binits{S.}},
\bauthor{\bsnm{{Nakwacki}}, \binits{M.S.}},
\bauthor{\bsnm{{D{\'e}moulin}}, \binits{P.}},
\bauthor{\bsnm{{Mandrini}}, \binits{C.H.}}:
\byear{2007},
\bjtitle{Solar Phys.}
\bvolume{244},
\bfpage{115}.
doi:\doiurl{10.1007/s11207-007-9034-2}.
\end{barticle}
\endbibitem

\bibitem[\protect\citeauthoryear{{Delaboudini{\`e}re}
  \textit{et~al.}}{1995}]{delaboudiniere:1995}
\begin{barticle}
\bauthor{\bsnm{{Delaboudini{\`e}re}}, \binits{J.}},
\bauthor{\bsnm{{Artzner}}, \binits{G.E.}},
\bauthor{\bsnm{{Brunaud}}, \binits{J.}},
\bauthor{\bsnm{{Gabriel}}, \binits{A.H.}},
\bauthor{\bsnm{{Hochedez}}, \binits{J.F.}},
\bauthor{\bsnm{{Millier}}, \binits{F.}},
\bauthor{\bsnm{{Song}}, \binits{X.Y.}},
\bauthor{\bsnm{{Au}}, \binits{B.}},
\bauthor{\bsnm{{Dere}}, \binits{K.P.}},
\bauthor{\bsnm{{Howard}}, \binits{R.A.}},
\bauthor{\bsnm{{Kreplin}}, \binits{R.}},
\bauthor{\bsnm{{Michels}}, \binits{D.J.}},
\bauthor{\bsnm{{Moses}}, \binits{J.D.}},
\bauthor{\bsnm{{Defise}}, \binits{J.M.}},
\bauthor{\bsnm{{Jamar}}, \binits{C.}},
\bauthor{\bsnm{{Rochus}}, \binits{P.}},
\bauthor{\bsnm{{Chauvineau}}, \binits{J.P.}},
\bauthor{\bsnm{{Marioge}}, \binits{J.P.}},
\bauthor{\bsnm{{Catura}}, \binits{R.C.}},
\bauthor{\bsnm{{Lemen}}, \binits{J.R.}},
\bauthor{\bsnm{{Shing}}, \binits{L.}},
\bauthor{\bsnm{{Stern}}, \binits{R.A.}},
\bauthor{\bsnm{{Gurman}}, \binits{J.B.}},
\bauthor{\bsnm{{Neupert}}, \binits{W.M.}},
\bauthor{\bsnm{{Maucherat}}, \binits{A.}},
\bauthor{\bsnm{{Clette}}, \binits{F.}},
\bauthor{\bsnm{{Cugnon}}, \binits{P.}},
\bauthor{\bsnm{{van Dessel}}, \binits{E.L.}}:
\byear{1995},
\bjtitle{Solar Phys.}
\bvolume{162},
\bfpage{291}.
doi:\doiurl{10.1007/BF00733432}.
\end{barticle}
\endbibitem

\bibitem[\protect\citeauthoryear{{Golub} \textit{et~al.}}{2007}]{golub:2007}
\begin{barticle}
\bauthor{\bsnm{{Golub}}, \binits{L.}},
\bauthor{\bsnm{{Deluca}}, \binits{E.}},
\bauthor{\bsnm{{Austin}}, \binits{G.}},
\bauthor{\bsnm{{Bookbinder}}, \binits{J.}},
\bauthor{\bsnm{{Caldwell}}, \binits{D.}},
\bauthor{\bsnm{{Cheimets}}, \binits{P.}},
\bauthor{\bsnm{{Cirtain}}, \binits{J.}},
\bauthor{\bsnm{{Cosmo}}, \binits{M.}},
\bauthor{\bsnm{{Reid}}, \binits{P.}},
\bauthor{\bsnm{{Sette}}, \binits{A.}},
\bauthor{\bsnm{{Weber}}, \binits{M.}},
\bauthor{\bsnm{{Sakao}}, \binits{T.}},
\bauthor{\bsnm{{Kano}}, \binits{R.}},
\bauthor{\bsnm{{Shibasaki}}, \binits{K.}},
\bauthor{\bsnm{{Hara}}, \binits{H.}},
\bauthor{\bsnm{{Tsuneta}}, \binits{S.}},
\bauthor{\bsnm{{Kumagai}}, \binits{K.}},
\bauthor{\bsnm{{Tamura}}, \binits{T.}},
\bauthor{\bsnm{{Shimojo}}, \binits{M.}},
\bauthor{\bsnm{{McCracken}}, \binits{J.}},
\bauthor{\bsnm{{Carpenter}}, \binits{J.}},
\bauthor{\bsnm{{Haight}}, \binits{H.}},
\bauthor{\bsnm{{Siler}}, \binits{R.}},
\bauthor{\bsnm{{Wright}}, \binits{E.}},
\bauthor{\bsnm{{Tucker}}, \binits{J.}},
\bauthor{\bsnm{{Rutledge}}, \binits{H.}},
\bauthor{\bsnm{{Barbera}}, \binits{M.}},
\bauthor{\bsnm{{Peres}}, \binits{G.}},
\bauthor{\bsnm{{Varisco}}, \binits{S.}}:
\byear{2007},
\bjtitle{Solar Phys.}
\bvolume{243},
\bfpage{63}.
doi:\doiurl{10.1007/s11207-007-0182-1}.
\end{barticle}
\endbibitem

\bibitem[\protect\citeauthoryear{{Green} \textit{et~al.}}{2007}]{green:2007}
\begin{barticle}
\bauthor{\bsnm{{Green}}, \binits{L.M.}},
\bauthor{\bsnm{{Kliem}}, \binits{B.}},
\bauthor{\bsnm{{T{\"o}r{\"o}k}}, \binits{T.}},
\bauthor{\bsnm{{van Driel-Gesztelyi}}, \binits{L.}},
\bauthor{\bsnm{{Attrill}}, \binits{G.D.R.}}:
\byear{2007},
\bjtitle{Solar Phys.}
\bvolume{246},
\bfpage{365}.
doi:\doiurl{10.1007/s11207-007-9061-z}.
\end{barticle}
\endbibitem

\bibitem[\protect\citeauthoryear{{Handy} \textit{et~al.}}{1999}]{handy:1999}
\begin{barticle}
\bauthor{\bsnm{{Handy}}, \binits{B.N.}},
\bauthor{\bsnm{{Acton}}, \binits{L.W.}},
\bauthor{\bsnm{{Kankelborg}}, \binits{C.C.}},
\bauthor{\bsnm{{Wolfson}}, \binits{C.J.}},
\bauthor{\bsnm{{Akin}}, \binits{D.J.}},
\bauthor{\bsnm{{Bruner}}, \binits{M.E.}},
\bauthor{\bsnm{{Caravalho}}, \binits{R.}},
\bauthor{\bsnm{{Catura}}, \binits{R.C.}},
\bauthor{\bsnm{{Chevalier}}, \binits{R.}},
\bauthor{\bsnm{{Duncan}}, \binits{D.W.}},
\bauthor{\bsnm{{Edwards}}, \binits{C.G.}},
\bauthor{\bsnm{{Feinstein}}, \binits{C.N.}},
\bauthor{\bsnm{{Freeland}}, \binits{S.L.}},
\bauthor{\bsnm{{Friedlaender}}, \binits{F.M.}},
\bauthor{\bsnm{{Hoffmann}}, \binits{C.H.}},
\bauthor{\bsnm{{Hurlburt}}, \binits{N.E.}},
\bauthor{\bsnm{{Jurcevich}}, \binits{B.K.}},
\bauthor{\bsnm{{Katz}}, \binits{N.L.}},
\bauthor{\bsnm{{Kelly}}, \binits{G.A.}},
\bauthor{\bsnm{{Lemen}}, \binits{J.R.}},
\bauthor{\bsnm{{Levay}}, \binits{M.}},
\bauthor{\bsnm{{Lindgren}}, \binits{R.W.}},
\bauthor{\bsnm{{Mathur}}, \binits{D.P.}},
\bauthor{\bsnm{{Meyer}}, \binits{S.B.}},
\bauthor{\bsnm{{Morrison}}, \binits{S.J.}},
\bauthor{\bsnm{{Morrison}}, \binits{M.D.}},
\bauthor{\bsnm{{Nightingale}}, \binits{R.W.}},
\bauthor{\bsnm{{Pope}}, \binits{T.P.}},
\bauthor{\bsnm{{Rehse}}, \binits{R.A.}},
\bauthor{\bsnm{{Schrijver}}, \binits{C.J.}},
\bauthor{\bsnm{{Shine}}, \binits{R.A.}},
\bauthor{\bsnm{{Shing}}, \binits{L.}},
\bauthor{\bsnm{{Strong}}, \binits{K.T.}},
\bauthor{\bsnm{{Tarbell}}, \binits{T.D.}},
\bauthor{\bsnm{{Title}}, \binits{A.M.}},
\bauthor{\bsnm{{Torgerson}}, \binits{D.D.}},
\bauthor{\bsnm{{Golub}}, \binits{L.}},
\bauthor{\bsnm{{Bookbinder}}, \binits{J.A.}},
\bauthor{\bsnm{{Caldwell}}, \binits{D.}},
\bauthor{\bsnm{{Cheimets}}, \binits{P.N.}},
\bauthor{\bsnm{{Davis}}, \binits{W.N.}},
\bauthor{\bsnm{{Deluca}}, \binits{E.E.}},
\bauthor{\bsnm{{McMullen}}, \binits{R.A.}},
\bauthor{\bsnm{{Warren}}, \binits{H.P.}},
\bauthor{\bsnm{{Amato}}, \binits{D.}},
\bauthor{\bsnm{{Fisher}}, \binits{R.}},
\bauthor{\bsnm{{Maldonado}}, \binits{H.}},
\bauthor{\bsnm{{Parkinson}}, \binits{C.}}:
\byear{1999},
\bjtitle{Solar Phys.}
\bvolume{187},
\bfpage{229}.
doi:\doiurl{10.1023/A:1005166902804}.
\end{barticle}
\endbibitem

\bibitem[\protect\citeauthoryear{{Harra} \textit{et~al.}}{2007}]{Harra&al2007}
\begin{barticle}
\bauthor{\bsnm{{Harra}}, \binits{L.K.}},
\bauthor{\bsnm{{Crooker}}, \binits{N.U.}},
\bauthor{\bsnm{{Mandrini}}, \binits{C.H.}},
\bauthor{\bsnm{{van Driel-Gesztelyi}}, \binits{L.}},
\bauthor{\bsnm{{Dasso}}, \binits{S.}},
\bauthor{\bsnm{{Wang}}, \binits{J.}},
\bauthor{\bsnm{{Elliott}}, \binits{H.}},
\bauthor{\bsnm{{Attrill}}, \binits{G.}},
\bauthor{\bsnm{{Jackson}}, \binits{B.V.}},
\bauthor{\bsnm{{Bisi}}, \binits{M.M.}}:
\byear{2007},
\bjtitle{Solar Phys.}
\bvolume{244},
\bfpage{95}.
doi:\doiurl{10.1007/s11207-007-9002-x}.
\end{barticle}
\endbibitem

\bibitem[\protect\citeauthoryear{Howard \textit{et~al.}}{2008}]{howard:2008}
\begin{barticle}
\bauthor{\bsnm{Howard}, \binits{R.A.}},
\bauthor{\bsnm{Moses}, \binits{J.D.}},
\bauthor{\bsnm{Vourlidas}, \binits{A.}},
\bauthor{\bsnm{Newmark}, \binits{J.S.}},
\bauthor{\bsnm{Socker}, \binits{D.G.}},
\bauthor{\bsnm{Plunkett}, \binits{S.P.}},
\bauthor{\bsnm{Korendyke}, \binits{C.M.}},
\bauthor{\bsnm{Cook}, \binits{J.W.}},
\bauthor{\bsnm{Hurley}, \binits{A.}},
\bauthor{\bsnm{Davila}, \binits{J.M.}},
\bauthor{\bsnm{Thompson}, \binits{W.T.}},
\bauthor{\bsnm{Cyr}, \binits{O.C.S.}},
\bauthor{\bsnm{Mentzell}, \binits{E.}},
\bauthor{\bsnm{Mehalick}, \binits{K.}},
\bauthor{\bsnm{Lemen}, \binits{J.R.}},
\bauthor{\bsnm{Wuelser}, \binits{J.P.}},
\bauthor{\bsnm{Duncan}, \binits{D.W.}},
\bauthor{\bsnm{Tarbell}, \binits{T.D.}},
\bauthor{\bsnm{Wolfson}, \binits{C.J.}},
\bauthor{\bsnm{Moore}, \binits{A.}},
\bauthor{\bsnm{Harrison}, \binits{R.A.}},
\bauthor{\bsnm{Waltham}, \binits{N.R.}},
\bauthor{\bsnm{Lang}, \binits{J.}},
\bauthor{\bsnm{Davis}, \binits{C.J.}},
\bauthor{\bsnm{Eyles}, \binits{C.J.}},
\bauthor{\bsnm{Mapson-Menard}, \binits{H.}},
\bauthor{\bsnm{Simnett}, \binits{G.M.}},
\bauthor{\bsnm{Halain}, \binits{J.P.}},
\bauthor{\bsnm{Defise}, \binits{J.M.}},
\bauthor{\bsnm{Mazy}, \binits{E.}},
\bauthor{\bsnm{Rochus}, \binits{P.}},
\bauthor{\bsnm{Mercier}, \binits{R.}},
\bauthor{\bsnm{Ravet}, \binits{M.F.}},
\bauthor{\bsnm{Delmotte}, \binits{F.}},
\bauthor{\bsnm{Auchere}, \binits{F.}},
\bauthor{\bsnm{Delaboudiniere}, \binits{J.P.}},
\bauthor{\bsnm{Bothmer}, \binits{V.}},
\bauthor{\bsnm{Deutsch}, \binits{W.}},
\bauthor{\bsnm{Wang}, \binits{D.}},
\bauthor{\bsnm{Rich}, \binits{N.}},
\bauthor{\bsnm{Cooper}, \binits{S.}},
\bauthor{\bsnm{Stephens}, \binits{V.}},
\bauthor{\bsnm{Maahs}, \binits{G.}},
\bauthor{\bsnm{Baugh}, \binits{R.}},
\bauthor{\bsnm{Mcmullin}, \binits{D.}}:
\byear{2008},
\bjtitle{Space Sci. Rev.}
\bvolume{136},
\bfpage{67}.
\end{barticle}
\endbibitem

\bibitem[\protect\citeauthoryear{Isenberg and Forbes}{2007}]{isenberg:2007}
\begin{barticle}
\bauthor{\bsnm{Isenberg}, \binits{P.A.}},
\bauthor{\bsnm{Forbes}, \binits{T.G.}}:
\byear{2007},
\bjtitle{Astrophys. J.}
\bvolume{670},
\bfpage{1453}.
\end{barticle}
\endbibitem

\bibitem[\protect\citeauthoryear{Kliem and T\"{o}r\"{o}k}{2006}]{kliem:2006}
\begin{barticle}
\bauthor{\bsnm{Kliem}, \binits{B.}},
\bauthor{\bsnm{T\"{o}r\"{o}k}, \binits{T.}}:
\byear{2006},
\bjtitle{Phys. Rev. Lett.}
\bvolume{96},
\bfpage{255002}.
\end{barticle}
\endbibitem

\bibitem[\protect\citeauthoryear{{Kliem}, {T\"{o}r\"{o}k}, and
  {Thompson}}{2011}]{Kliem&al2011}
\begin{botherref}
\oauthor{\bsnm{{Kliem}}, \binits{B.}},
\oauthor{\bsnm{{T\"{o}r\"{o}k}}, \binits{T.}},
\oauthor{\bsnm{{Thompson}}, \binits{W.T.}}:
2011,
\textit{Solar Phys., to be submitted (Paper~II)}.
\end{botherref}
\endbibitem

\bibitem[\protect\citeauthoryear{{Ko} \textit{et~al.}}{2010}]{ko:2010}
\begin{barticle}
\bauthor{\bsnm{{Ko}}, \binits{Y.}},
\bauthor{\bsnm{{Raymond}}, \binits{J.C.}},
\bauthor{\bsnm{{Vrsnak}}, \binits{B.}},
\bauthor{\bsnm{{Vujic}}, \binits{E.}}:
\byear{2010},
\bjtitle{Astrophys. J.}
\bvolume{722},
\bfpage{625}.
doi:\doiurl{10.1088/0004-637X/722/1/625}.
\end{barticle}
\endbibitem

\bibitem[\protect\citeauthoryear{{Landi} \textit{et~al.}}{2010}]{Landi&al2010}
\begin{barticle}
\bauthor{\bsnm{{Landi}}, \binits{E.}},
\bauthor{\bsnm{{Raymond}}, \binits{J.C.}},
\bauthor{\bsnm{{Miralles}}, \binits{M.P.}},
\bauthor{\bsnm{{Hara}}, \binits{H.}}:
\byear{2010},
\bjtitle{Astrophys. J.}
\bvolume{711},
\bfpage{75}.
doi:\doiurl{10.1088/0004-637X/711/1/75}.
\end{barticle}
\endbibitem

\bibitem[\protect\citeauthoryear{{Martin}}{1998}]{martin:1998}
\begin{barticle}
\bauthor{\bsnm{{Martin}}, \binits{S.F.}}:
\byear{1998},
\bjtitle{Solar Phys.}
\bvolume{182},
\bfpage{107}.
doi:\doiurl{10.1023/A:1005026814076}.
\end{barticle}
\endbibitem

\bibitem[\protect\citeauthoryear{{Martin}}{2003}]{martin:2003}
\begin{barticle}
\bauthor{\bsnm{{Martin}}, \binits{S.F.}}:
\byear{2003},
\bjtitle{Adv. Space Res.}
\bvolume{32},
\bfpage{1883}.
doi:\doiurl{10.1016/S0273-1177(03)90622-3}.
\end{barticle}
\endbibitem

\bibitem[\protect\citeauthoryear{Martin and McAllister}{1997}]{martin:1997}
\begin{bchapter}
\bauthor{\bsnm{Martin}, \binits{S.F.}},
\bauthor{\bsnm{McAllister}, \binits{A.H.}}:
\byear{1997},
In: \beditor{\bsnm{Crooker}, \binits{N.}},
\beditor{\bsnm{Joselyn}, \binits{J.A.}},
\beditor{\bsnm{Feynman}, \binits{J.}} (eds.)
\bbtitle{{\it Coronal Mass Ejections: Causes and Consequences}, Geophys.
  Monogr. Ser. 99},
\bpublisher{AGU},
\blocation{Washington, D.C.},
\bfpage{127}.
\end{bchapter}
\endbibitem

\bibitem[\protect\citeauthoryear{{Martin}, {Bilimoria}, and
  {Tracadas}}{1994}]{martin:1994}
\begin{bchapter}
\bauthor{\bsnm{{Martin}}, \binits{S.F.}},
\bauthor{\bsnm{{Bilimoria}}, \binits{R.}},
\bauthor{\bsnm{{Tracadas}}, \binits{P.W.}}:
\byear{1994},
In: \beditor{\bsnm{{R.~J.~Rutten \& C.~J.~Schrijver}}} (ed.)
\bbtitle{Solar Surface Magnetism},
\bfpage{303}.
\end{bchapter}
\endbibitem

\bibitem[\protect\citeauthoryear{{Muglach}, {Wang}, and
  {Kliem}}{2009}]{Muglach&al2009}
\begin{barticle}
\bauthor{\bsnm{{Muglach}}, \binits{K.}},
\bauthor{\bsnm{{Wang}}, \binits{Y.}},
\bauthor{\bsnm{{Kliem}}, \binits{B.}}:
\byear{2009},
\bjtitle{Astrophys. J.}
\bvolume{703},
\bfpage{976}.
doi:\doiurl{10.1088/0004-637X/703/1/976}.
\end{barticle}
\endbibitem

\bibitem[\protect\citeauthoryear{{Patsourakos} and
  {Vourlidas}}{2011}]{Patsourakos&Vourlidas2011}
\begin{barticle}
\bauthor{\bsnm{{Patsourakos}}, \binits{S.}},
\bauthor{\bsnm{{Vourlidas}}, \binits{A.}}:
\byear{2011},
\bjtitle{Astron. Astrophys.}
\bvolume{525},
\bfpage{A27+}.
doi:\doiurl{10.1051/0004-6361/201015048}.
\end{barticle}
\endbibitem

\bibitem[\protect\citeauthoryear{Pevtsov, Balasubramaniam, and
  Rogers}{2003}]{pevtsov:2003}
\begin{barticle}
\bauthor{\bsnm{Pevtsov}, \binits{A.A.}},
\bauthor{\bsnm{Balasubramaniam}, \binits{K.S.}},
\bauthor{\bsnm{Rogers}, \binits{J.W.}}:
\byear{2003},
\bjtitle{Astrophys. J.}
\bvolume{595},
\bfpage{500}.
\end{barticle}
\endbibitem

\bibitem[\protect\citeauthoryear{{Plunkett}
  \textit{et~al.}}{2000}]{Plunkett&al2000}
\begin{barticle}
\bauthor{\bsnm{{Plunkett}}, \binits{S.P.}},
\bauthor{\bsnm{{Vourlidas}}, \binits{A.}},
\bauthor{\bsnm{{{\v S}imberov{\'a}}}, \binits{S.}},
\bauthor{\bsnm{{Karlick{\'y}}}, \binits{M.}},
\bauthor{\bsnm{{Kotr{\v c}}}, \binits{P.}},
\bauthor{\bsnm{{Heinzel}}, \binits{P.}},
\bauthor{\bsnm{{Kupryakov}}, \binits{Y.A.}},
\bauthor{\bsnm{{Guo}}, \binits{W.P.}},
\bauthor{\bsnm{{Wu}}, \binits{S.T.}}:
\byear{2000},
\bjtitle{Solar Phys.}
\bvolume{194},
\bfpage{371}.
\end{barticle}
\endbibitem

\bibitem[\protect\citeauthoryear{{Priest} and
  {Forbes}}{2002}]{Priest&Forbes2002}
\begin{barticle}
\bauthor{\bsnm{{Priest}}, \binits{E.R.}},
\bauthor{\bsnm{{Forbes}}, \binits{T.G.}}:
\byear{2002},
\bjtitle{Astron.\ Astrophys.\ Rev.}
\bvolume{10},
\bfpage{313}.
doi:\doiurl{10.1007/s001590100013}.
\end{barticle}
\endbibitem

\bibitem[\protect\citeauthoryear{Rust and La{B}onte}{2005}]{rust:2005}
\begin{barticle}
\bauthor{\bsnm{Rust}, \binits{D.M.}},
\bauthor{\bsnm{La{B}onte}, \binits{B.J.}}:
\byear{2005},
\bjtitle{Astrophys. J.}
\bvolume{622},
\bfpage{69}.
\end{barticle}
\endbibitem

\bibitem[\protect\citeauthoryear{{Ruzmaikin}, {Martin}, and
  {Hu}}{2003}]{ruzmaikin:2003}
\begin{barticle}
\bauthor{\bsnm{{Ruzmaikin}}, \binits{A.}},
\bauthor{\bsnm{{Martin}}, \binits{S.}},
\bauthor{\bsnm{{Hu}}, \binits{Q.}}:
\byear{2003},
\bjtitle{J. Geophys. Res.}
\bvolume{108},
\bfpage{1096}.
doi:\doiurl{10.1029/2002JA009588}.
\end{barticle}
\endbibitem

\bibitem[\protect\citeauthoryear{{Savage} \textit{et~al.}}{2010}]{savage:2010}
\begin{barticle}
\bauthor{\bsnm{{Savage}}, \binits{S.L.}},
\bauthor{\bsnm{{McKenzie}}, \binits{D.E.}},
\bauthor{\bsnm{{Reeves}}, \binits{K.K.}},
\bauthor{\bsnm{{Forbes}}, \binits{T.G.}},
\bauthor{\bsnm{{Longcope}}, \binits{D.W.}}:
\byear{2010},
\bjtitle{Astrophys. J.}
\bvolume{722},
\bfpage{329}.
doi:\doiurl{10.1088/0004-637X/722/1/329}.
\end{barticle}
\endbibitem

\bibitem[\protect\citeauthoryear{T\"{o}r\"{o}k, Kliem, and
  Titov}{2004}]{torok:2004}
\begin{barticle}
\bauthor{\bsnm{T\"{o}r\"{o}k}, \binits{T.}},
\bauthor{\bsnm{Kliem}, \binits{B.}},
\bauthor{\bsnm{Titov}, \binits{V.S.}}:
\byear{2004},
\bjtitle{Astron. Astrophys.}
\bvolume{413},
\bfpage{27}.
\end{barticle}
\endbibitem

\bibitem[\protect\citeauthoryear{{Wang}, {Muglach}, and
  {Kliem}}{2009}]{wang:2009}
\begin{barticle}
\bauthor{\bsnm{{Wang}}, \binits{Y.}},
\bauthor{\bsnm{{Muglach}}, \binits{K.}},
\bauthor{\bsnm{{Kliem}}, \binits{B.}}:
\byear{2009},
\bjtitle{Astrophys. J.}
\bvolume{699},
\bfpage{133}.
doi:\doiurl{10.1088/0004-637X/699/1/133}.
\end{barticle}
\endbibitem

\bibitem[\protect\citeauthoryear{{Yurchyshyn}}{2008}]{Yurchyshyn2008}
\begin{barticle}
\bauthor{\bsnm{{Yurchyshyn}}, \binits{V.}}:
\byear{2008},
\bjtitle{Astrophys. J.}
\bvolume{675},
\bfpage{49}.
doi:\doiurl{10.1086/533413}.
\end{barticle}
\endbibitem

\bibitem[\protect\citeauthoryear{{Yurchyshyn}, {Abramenko}, and
  {Tripathi}}{2009}]{Yurchyshyn&al2009}
\begin{barticle}
\bauthor{\bsnm{{Yurchyshyn}}, \binits{V.}},
\bauthor{\bsnm{{Abramenko}}, \binits{V.}},
\bauthor{\bsnm{{Tripathi}}, \binits{D.}}:
\byear{2009},
\bjtitle{Astrophys. J.}
\bvolume{705},
\bfpage{426}.
doi:\doiurl{10.1088/0004-637X/705/1/426}.
\end{barticle}
\endbibitem

\end{thebibliography}

\end{article}
\end{document}